\documentclass[useAMS,usenatbib]{mnras}

\usepackage[utf8]{inputenc}
\usepackage{amsmath,amsfonts,amssymb,slashed}
\usepackage{graphicx}
\usepackage{nicefrac}
\usepackage{tabularx}

\title{Outliers in the LIGO Black Hole Mass Function from Coagulation in Dense Clusters}\author[Jordan Flitter et al.]{Jordan Flitter,$^{1,2}$
Julian B.\ Mu\~noz,$^{3,4}$,
and Ely D. Kovetz,$^{2}$ \\
$^{1}$ Physics Department, Tel-Aviv University, Ramat-Aviv 69978, Israel\\
$^{2}$ Department of Physics, Ben-Gurion University of the Negev, Be'er Sheva 84105, Israel\\
$^{3}$ Department of Physics, Harvard University, Cambridge, MA 02138, USA\\
$^{4}$ Harvard-Smithsonian Center for Astrophysics, Cambridge, MA 02138, USA\\
}

\begin{document}
\label{firstpage}
\pagerange{\pageref{firstpage}--\pageref{lastpage}}
\maketitle

\begin{abstract}
The advanced LIGO O3a run catalog has been recently published, and it includes several events with unexpected mass properties, including mergers with individual masses in the lower and upper mass gaps, as well as mergers with unusually small mass ratios between the binary components.
Here we entertain the possibility that these outliers are the outcome of hierarchical mergers of black holes or neutron stars in the dense environments of globular clusters. We use the coagulation equation to study the evolution of the black hole mass function within a typical cluster. Our prescription allows us to monitor how various global quantities change with time, such as the total mass and number of compact objects in the cluster,  its overall merger rate, and the probability to form intermediate-mass black holes via a runaway process. By accounting for the LIGO observational bias, we predict the merger event distributions with respect to various variables such as the individual masses $M_1$ and $M_2$, their ratio $q$, and redshift $z$, and we compare our predictions with the published O3a data. We study how these distributions depend on the merger-rate and ejections parameters and produce forecasts for the (tight) constraints that can be placed on our model parameters using the future dataset of the O5 run. Finally, we also consider the presence of a static channel with no coagulation producing merger events alongside the dynamic channel, finding that the two can be distinguished based solely on the merger mass distribution with future O5 data. 
\end{abstract}

\begin{keywords}
black holes, gravitational waves 
\end{keywords}

\section{Introduction}

It was only in early 2016 when the Laser Interferometer Gravitational-Wave Observatory (LIGO) collaboration announced the first detection of a gravitational wave (GW) signal, known as GW150914~\citep{Abbott:2016blz}. The estimated masses of the two black holes (BHs) in the merger were $\gtrsim30M_\odot$. This discovery did not only confirm Einstein's theory of general relativity, but also heralded the era of BH observation via the detection of GWs that are produced in their coalescence. 
Since the discovery of GW150914, nine more BH coalescence events were discovered during LIGO's first two runs, and their properties were published in the GWTC-1 catalog~\citep{LIGOScientific:2018mvr}.

The inferred BH masses of the GWTC-1 events indicated that BH mergers tend to happen between components with a mass ratio of  $q\sim1$, and that the amount of detected events in the region $30M_\odot \lesssim M$ is roughly twice the amount of detected events in the region $M\lesssim30M_\odot$. This statement does not necessarily imply that there are more heavy BHs than lighter ones, however, since LIGO is more sensitive to mergers of the former. This observational bias changes the observed BH mass function (BHMF) relative to the true BHMF~\citep{Kovetz:2016kpi}.

The third run of advanced LIGO (aLIGO), O3, ended in March 2020. By the end of October 2020 the properties of the 39 events that were detected during the first six months of O3 (O3a) were published in the GWTC-2 catalog~\citep{Abbott:2020niy,Abbott:2020gyp}. While most of the properties of O3a events agree with the properties of the known BHs from GWTC-1, attention has been focused on several outlier events. First, the GW190814 event~\citep{Abbott:2020khf} includes an object on the verge of the lower mass gap --- the region in the mass spectrum bounded from below by $\sim3M_\odot$ (the theoretical limit for the maximal mass of a neutron star~\citep{Oppenheimer:1939ne}), and from above by $\sim5M_\odot$ (the empirical minimal mass of BHs that have been detected via X-ray binaries~\citep{Ozel:2010, Farr:2010tu}). Secondly, GWTC-2 contains $\gtrsim5$ events with BHs heavier than $50M_\odot$, including the famous GW190521 event~\citep{Abbott:2020tfl}. This implies a breach of the upper mass gap --- a dearth in BH of masses $M\gtrsim50\,M_{\odot}$ which is motivated by the pair-instability supernovae (PISNe) theory~\citep{Talbot:2018cva, Belczynski:2016jno, Stevenson:2019rcw, Woosley:2016hmi}. Thirdly, the GW190412 and GW190814 events~\citep{LIGOScientific:2020stg, Abbott:2020khf} reveal that BH mergers with small mass ratios ($q<0.5$) are more common than originally thought.

Motivated by these outliers, in this paper we study the outcome of hierarchical dynamical mergers of BHs and neutron stars (NSs) in globular clusters (GCs). We use an analytical tool, the coagulation equation~\citep{Smol,Mouri:2002mc}, to study the evolution of the BHMF within the typical GC. \cite{Christian:2018mjv} had solved the continuous form of the coagulation equation in that context for different scenarios where either the {\it overall} merger rate, the {\it total fraction} of ejections, or the interaction kernel were altered, but without considering the observational bias. Meanwhile, using monte-carlo simulations with a discrete-step promotion algorithm that only mimics the properties of the coagulation equation, ~\cite{Doctor:2019ruh} obtained the cluster BHMF as well as the predicted mass distribution of detected events for a given overall merger rate and interaction kernel. 

Here we introduce a physics-driven model for the ejections, mass loss and delay time during mergers and solve the discrete form of the coagulation equation. Crucially, our prescription allows us to monitor how global features of the GC change with time --- such as the total number of BHs, the total BHs mass, the total merger rate, and the probability to form intermediate-mass black holes (IMBHs) --- and to consistently evolve the merger rate according to the cluster evolution (rather than setting it constant~\citep{Christian:2018mjv,Doctor:2019ruh}).  We account for the aLIGO observational bias and predict the observed distribution of merger events with respect to various source variables such as $M_1$, $M_2$, $q$ and $z$. 

This consistent analytical framework we construct is the primary result of this paper. As we demonstrate below, it can be easily used to combine multiple channels with different properties~\citep{Rodriguez:2016vmx,Zevin:2017evb,Inayoshi:2017mrs,Bird:2016dcv,Kovetz:2017rvv,Liu:2020ufc,Kritos:2020fjw} and to incorporate more sophisticated modeling of various elements such as GC density profiles, velocity dispersions, natal kicks, ejection mechanisms, etc. It can also be used to fit the model parameters to the data using standard bayesian analysis.

Here, with the observed O3a distributions of individual masses and the mass ratio of detected events in mind, we compare and contrast our results for the dynamical (GC) channel with those from a static (field) channel,  where the BH mass function from which merging BH are drawn does not evolve with redshift. Several qualitative conclusions can be gleaned from this comparison:

First, our model naturally suggests that breaching the upper mass gap via coagulation in dense stellar environments is achievable. Secondly, we demonstrate that under certain circumstances, coagulation of NSs can form BHs in the lower mass gap which can be detected with LIGO O3 sensitivity (and hence it is plausible, though not necessarily likely\footnote{A thorough examination of this scenario is left to future work.}, that the peculiar object of mass $2.6M_\odot$ that was detected in GW190814 is a 2nd-generation BH resulting from NS coagulation in GCs). 

Moreover, our results indicate that consistency with GWTC-2 requires that the interaction rate kernel of both the dynamic and the static channel prefers equal-mass mergers. Under this assumption, we also find that the probability that an IMBH\footnote{We define BHs with mass $>500\,M_\odot$ as IMBHs. Although in the literature there is an upper bound of order $\sim10^6M_\odot$ for a BH to be considered as an IMBH, above which it is considered as a super-massive BH (SMBH), in this work we never entertain GCs with such high total mass values and therefore we consider an IMBH to be any BH with mass above $500M_\odot$.} is formed via hierarchical mergers of SBHs is exceedingly small within a Hubble time, suggesting that IMBHs are formed via other mechanisms ~\citep{PflammAltenburg:2009ij, KSJW:2020}.
Finally, using Fisher analysis we forecast  that tight constraints can be placed on the free model parameters given future aLIGO O5 sensitivity. Importantly, we find that the expected rich dataset of O5 can determine (at $>2\sigma$ CL) whether the detected mergers originated from both  dynamic and static environments.

This paper is constructed as follows: Section \ref{The coagulation equation} provides a mathematical background to the application of the coagulation equation in GC environments and the subtleties that  need to be taken care of when solving this equation numerically. Section \ref{Our Model} gives a full description of our GC model. We show examples for possible solutions to the coagulation equation in the context of our GC model in section \ref{Coagulation results}. In section \ref{The observed merger density function} we introduce a second static channel and describe our prescription for calculating the merger events distribution with respect to the mass components and the event redshifts. We discuss the observable features of the dynamic channel and the O3a outliers in section \ref{Comparison to observations} and associate these outliers with our model parameters. In section \ref{Forecasts} we combine the contributions from the static and dynamic channels and forecast constraints on our model parameters  considering the future O5 run. We conclude in section \ref{Conclusions}.

\section{The coagulation equation}\label{The coagulation equation}

The function that describes the probability of a BH in a certain environment to have a mass $M$ at time $t$ is called the black hole mass function (BHMF), which we denote by $f\left(M,t\right)$. We assume that the initial mass function (IMF) is inherited from that of stars~\citep{Salpeter:1955it}, and its evolution then depends on the dynamics of BHs in each environment.

Here we will study the evolution of $f\left(M,t\right)$ in the dense environments of globular clusters (GCs), where successive BHs mergers are possible, through the coagulation (or \cite{Smol}) equation. 
We assume that the masses of all the BHs in the GC are integer multiples of a mass quanta $M_0$ (for computational purposes only, as $M$ is a continuous quantity) . This means that each BH mass in the GC, at any given time, can be written as $M=M_{i}\equiv i\, M_{0}$, where $i$ is a positive integer. We use the notation $N_i\left(t\right)$ to denote the number of BHs of mass $M_i$ in the GC at time $t$. For a small-enough value of $M_0$ (as we will describe later) there is a direct relation between $N_i\left(t\right)$ and the BHMF, namely $N_{i}\left(t\right)=M_{0}f\left(M_{i},t\right)$. We will denote $\Gamma_{i,j}$ as the interaction rate between BHs of masses $M_{i}$ and $M_{j}$ in the GC. 
With this notation, the discrete form of the coagulation equation is given by
\begin{equation}\label{Discrete coagulation}
\dot{N}_{i}=\frac{1}{2}\sum_{j=1}^{i-1}\Gamma_{i-j,j}-\sum_{j=1}^{\infty}\Gamma_{i,j}.
\end{equation}
The LHS of Eq.~(\ref{Discrete coagulation}) represents the change in $N_{i}$. On the one hand, if smaller BHs merge to form a new BH of mass $M_{i}$, then $N_{i}$ increases, which is reflected by the first term in the RHS\footnote{The factor half is required since the summation counts every possible interaction twice.}. On the other hand, if a BH of mass $M_{i}$ merges with any other BH, then $N_{i}$ decreases, as governed by the second term in the RHS.

In order to account for various effects in the evolution of the GC, the coagulation equation has to be modified. We consider four such enhancements:
\begin{enumerate}
\item\textbf{Finite total mass.} To avoid mergers that form a BH with mass greater than $M_\mathrm{tot}$, the total BH mass, we truncate the second sum at  $j_\mathrm{max}=i_\mathrm{max}-i$ where $i_{\mathrm{max}}\left(t\right)\equiv M_{\mathrm{tot}}\left(t\right)/M_0=\sum_{i=1}^{\infty}iN_{i}\left(t\right)$.

\item\textbf{Mass loss.} During the coalescence of BHs, the radiated GWs carry energy away from the binary BH (BBH) system. Hence, mass is not conserved and the mass of the remnant BH is not simply the sum of the coalescing BHs, but $M_{\mathrm{rem}}\left(M_j,M_k\right)=M_j+M_k-E_{\mathrm{GW}}\left(M_j,M_k\right)$, where $E_{\mathrm{GW}}$ is the total energy of the GWs that were produced in the coalescence. To account for this effect, the first term in Eq.~(\ref{Discrete coagulation}) is turned into  a double sum  restricted by the Kronecker delta $\delta_{i,i_{\mathrm{rem}}}$, where $i_{\mathrm{rem}}\left(j,k\right)=M_{\mathrm{rem}}/M_0$. Under this modification, $j_{\mathrm{max}}\left(i,i_{\mathrm{max}}\right)$ is the maximal solution to the remnant mass equation when $i_{\mathrm{rem}}$ is replaced with $i_{\mathrm{max}}$.\footnote{Which is $i_\mathrm{max}-i$ if $E_{\mathrm{GW}}\equiv0$.}

\item\textbf{Ejections.} It is fairly possible that BHs will be ejected from the GC during its evolution. There are two types of ejections that we can consider. The first is due to asymmetry in the emission of GWs in BH mergers. GWs carry linear momentum, and when the coalescing masses are not equal, or their spins are misaligned with the orbital axis, the total linear momentum of the GWs can be non-isotropic, resulting in a recoil to the BHs ~\citep{f83, Hughes:2004ck, HolleyBockelmann:2007eh, Fragione:2017blf, Merritt:2004xa, Favata:2004wz, Gerosa:2019zmo, Varma:2020nbm}. The second type of ejections is due to 3-body interactions \citep{Samsing:2017xmd, Samsing:2017plz, Samsing:2019dtb, Samsing:2020qqd}. As a chaotic system, there are many possible outcomes for 3-body interactions, which depend upon the exact initial configuration of the system. One configuration which is commonly studied is a BBH and a single disrupting BH. The single BH may pair up with one of the binary members, resulting in a kick to the second member through a sling-shot interaction. The recoil velocity of the BH can be large enough that it is often ejected from the cluster. In this work we consider the two types of ejections.
We model the ejections by introducing an ejection rate term $\Gamma_\mathrm{ej}\left(M_i,t\right)$ to the RHS of Eq.~(\ref{Discrete coagulation}). This $\Gamma_{\mathrm{ej}}$ is expected to be the sum (with respect to $j$) of the interaction rate $\Gamma_{i,j}$ times the probability\footnote{ $P_\mathrm{ej}\left(M_i,M_j\right)$ may depend on the spins of the BHs, see details underneath Eq.~\eqref{Probability for ejection}.} to have an ejection $P_\mathrm{ej}\left(M_i,M_j\right)$. This method for modelling ejections using their probability is similar to what was considered in \cite{Christian:2018mjv}, however in our analysis we derive the dependence of $P_\mathrm{ej}$ on the recoil velocity - see appendix \ref{Derivation for the relation between the probability for ejection and the recoil velocity}.

\item\textbf{Delay time.} Eq.~(\ref{Discrete coagulation}) assumes that once an interaction between two BHs has occurred, the BHMF is changed instantaneously. However, $\Gamma_{i,j}$ is not the merger rate --- it is the interaction rate to from BH binaries. Thus, in order to account for the delay between the formation time and the merger time, we must convolve the interaction rate with a delay time distribution $P\left(t_d\right)$, namely $\tilde\Gamma_{i,j}\left(t\right)=\int_0^t\Gamma_{i,j}\left(t-t_d\right)P\left(t_d\right)dt_d$, where $\tilde\Gamma_{i,j}\left(t\right)$ is the merger rate at time $t$. We assume that a possible ejection takes place immediately after the binary has formed, implying that the ejection rate is not altered due to the delay time between formation and merger. For that reason, we do not perform any convolution with the ejection term in the coagulation equation.

\end{enumerate}

\begin{figure}
\includegraphics[width=0.4\textwidth]{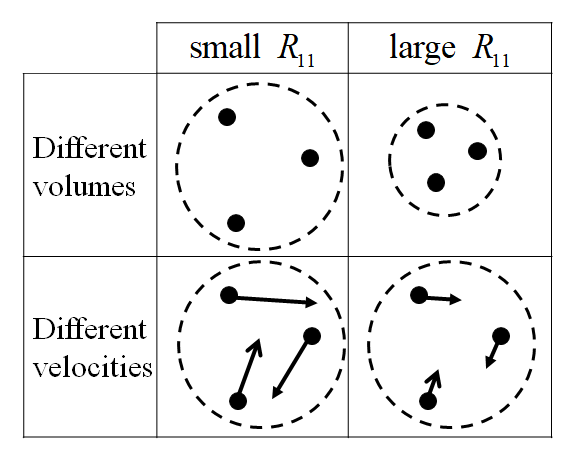}
\caption{Illustration: Dependence of the rate $R_{11}$ on cluster properties. Higher density or smaller  velocity dispersion yield larger $R_{11}$.}
\label{figure_1}
\end{figure}

Under these modifications, the coagulation equation becomes
\begin{equation}\label{coagulation equation}
\dot{N}_{i}=\frac{1}{2}\sum_{j=1}^{i_{\mathrm{max}}}\sum_{k=1}^{i_{\mathrm{max}}}\tilde\Gamma_{j,k}\delta_{i,i_{\mathrm{rem}}}-\sum_{j=1}^{j_{\mathrm{max}}}\left[\tilde\Gamma_{i,j}+\Gamma_{i,j}P_\mathrm{ej}\left(M_i,M_j\right)\right].
\end{equation}
Since the interaction rate $\Gamma_{i,j}$ is proportional to the number of BHs of masses $M_{i}$ and $M_{j}$, a more fundamental quantity to consider is the normalized interaction rate $R_{i,j}$, defined as
\begin{equation}\label{Gamma_i,j}
\Gamma_{i,j}\left(t\right)=R_{i,j}N_i\left(t\right)N_j\left(t\right)=R_{11}\tilde{R}_{i,j}N_i\left(t\right)N_j\left(t\right),
\end{equation}
where $\tilde{R}_{i,j}$ is the interaction rate kernel, which encompasses the physical properties of each individual interaction, i.e.\ how likely it is for masses $M_i$ and $M_j$ to interact. $R_{11}$ serves as an overall normalization factor to the interaction rate of the cluster: denser clusters or clusters with smaller dispersion velocity have greater $R_{11}$ as demonstrated in Fig.~\ref{figure_1}. An example for the expression of $R_{11}$ in the case of two-body gravitational capture processes is given in appendix \ref{Derivation for 2-body capture rate kernel}.

\section{The Model}\label{Our Model}

In this work we will solve the coagulation equation for a specific GC which is assumed to be the average GC whose mergers are observed by aLIGO. We use Eq.~(\ref{coagulation equation}) to promote the BHMF of a GC from $t=0$ to $t_\mathrm{c}=13\,\mathrm{Gyrs}$. The only elements that  need to be specified in order to solve the coagulation equation are the IMF $N_i\left(t=0\right)$, the interaction rate kernel $\tilde{R}_{i,j}$, the overall interaction rate regulator $R_{11}$, the delay time distribution $P\left(t_d\right)$, the amount of energy carried away by GWs, $E_{\mathrm{GW}}\left(M_i,M_j\right)$, and the ejection probability model $P_{\mathrm{ej}}\left(M_i,M_j\right)$. In the following subsections we describe our model for each of these elements. For convenience, all the parameters of our model are listed in table \ref{table_1}.

\begin{center}
\begin{table}
\begin{tabularx}{0.48\textwidth}{|X<{\setlength\hsize{0.8\hsize}\centering} | X<{\setlength\hsize{1.2\hsize}\centering}|}
\hline Parameters & Meaning \\ \hline\hline
  $\alpha$, $M_\mathrm{min}$ ,$M_\mathrm{max}$ &  BHMF parameters \\ \hline
  $\mu_\mathrm{pp}, \sigma_\mathrm{pp}$, $\lambda$ & PPSN parameters \\ \hline
  $\mu_\mathrm{NS}$, $\sigma_\mathrm{NS}$ & NS mass function parameters \\ \hline
  $N_\mathrm{0,SBH}$, $N_\mathrm{0,NS}$ & Initial number of objects \\ \hline
  $t_c$ & Total coagulation time \\ \hline
  $t_{d,\mathrm{min}}$ & Minimum delay time \\ \hline
  $R_{11}$, $R_0$ & Overall merger rate regulators \\ \hline
  $\beta$, $\beta_2$, $\gamma$ & Rate kernel parameters \\ \hline
  $f_\mathrm{loss}$ & GW mass loss coefficient \\ \hline
  $M_\mathrm{esc}$, $v_\mathrm{esc}$ & Ejections parameters \\ \hline
  $X$ & Populations mixture parameter \\ \hline
\end{tabularx}
\caption{All the parameters of our model.}
\label{table_1}
\end{table}
\end{center}

\subsection{The IMF}\label{The IMF}
The IMF serves as an initial condition for the coagulation equation. Our model IMF has contributions from both stellar BHs (SBHs) and NSs. We follow \cite{Christian:2018mjv} and choose to model the SBH IMF using the function
\begin{equation}\label{Ely's IMF}
f_{0,\alpha}\left(M\right)\propto M^{-\alpha}\mathcal{H}\left(M-M_{\mathrm{min}}\right)\mathcal{H}\left(M_{\mathrm{max}}-M\right),
\end{equation}
where $\alpha$ is the Salpeter power-law index \citep{Salpeter:1955it}, $\mathcal{H}$ is the Heaviside function, $M_{\mathrm{min}}$ is a sharp cutoff located at the upper bound of the lower mass gap, and $M_{\mathrm{max}}$ is the pair-instability supernova (PISN) upper cutoff on the SBH mass \citep{Talbot:2018cva, Belczynski:2016jno, Stevenson:2019rcw, Woosley:2016hmi}. PISNe happen when the pressure of the dying stellar core is substantially reduced due to the production of electron-positron pairs. Stars above $\sim150M_\odot$ are expected to undergo such a runaway process, and terminate while leaving no remnant behind. This  predicts a dearth of BHs above $\sim50\,M_\odot$~\citep{Talbot:2018cva}. We therefore set $\alpha=2.35$ to match the Kroupa mass function \citep{Kroupa:2000iv} and take fiducial values of $M_{\mathrm{min}}=5M_{\odot}$ and $M_{\mathrm{max}}=50M_{\odot}$.

On top of this, we add the contribution of SBHs that originate from pulsational pair-instability supernovae (PPSNe). Stars with mass in the range $100\,M_\odot\lesssim M\lesssim150\,M_\odot$ undergo a series of explosions  in which large amounts of matter are ejected prior to the eventual collapse. The expected result of this process is that all such stars form SBHs with masses $\sim40\,M_\odot$. We follow \cite{Talbot:2018cva} and model the PPSN contribution to the IMF as a Gaussian with mean $\mu_{\mathrm{pp}}$ and variance $\sigma_{\mathrm{pp}}^{2}$. The SBH IMF is then a linear combination of two contributions,
\begin{equation}\label{stellar IMF}
f_{0,\mathrm{SBH}}\left(M\right)=\left(1-\lambda\right)f_{0,\alpha}\left(M\right)+\lambda f_{0,\mathrm{PPSN}}\left(M\right), 
\end{equation}
where $\lambda$ represents  the fractional contribution of PPSNe to the IMF. We adopt $\mu_{\mathrm{pp}}=35M_\odot$ and $\sigma_{\mathrm{pp}}=3M_\odot$ for the Gaussian parameters. As for $\lambda$, we assign this parameter the value $0.08$, which is consistent with the relative amount of stars in the region $100M_\odot\lesssim M\lesssim150M_\odot$ (assuming a Salpeter power-law of $\alpha=2.35$).

We will also consider mergers in GCs that host both a BH and NS population. 
We model the NS IMF as a Gaussian of mean $\mu_\mathrm{NS}=1.33M_\odot$ and standard deviation $\sigma_\mathrm{NS}=0.09M_\odot$, as is common (see \cite{Ozel:2016oaf,Farrow:2019xnc, Gupta:2019nwj})\footnote{We note that there is an uncertainty about the exact shape of the NS IMF and there are various competing models in the literature. However, the O3a sensitivity is not enough to distinguish between these models and thus our conclusions regarding the O3a run are not affected by the choice of the NS IMF. We leave to future work a full examination of NS IMF models.}. 
We further assume that each NS merger results in a BH \citep{Gupta:2019nwj}, and allow for these second-generation BHs to merge with other NSs or BHs in the cluster. 
The total IMF is a mixture between the SBHs and NSs populations,\footnote{$f_{0,\mathrm{SBH}}\left(M\right)$ and $f_{0,\mathrm{NS}}\left(M\right)$ are assumed to be normalized to sum to 1 given an initial number of compact objects $N_\mathrm{0,SBH}+N_\mathrm{0,NS}$.}
\begin{flalign}\label{SBH+NS IMF}
f_0\left(M\right)=N_\mathrm{0,SBH}\cdot f_{0,\mathrm{SBH}}\left(M\right)+N_\mathrm{0,NS}\cdot f_{0,\mathrm{NS}}\left(M\right),
\end{flalign}
where $N_\mathrm{0,SBH}$ and $N_\mathrm{0,NS}$ are the initial number of SBHs, NSs, respectively. We discretize the mass spectrum by using $N_i\left(t=0\right)=M_0f_0\left(M_i\right)$. We take $M_0=1M_\odot$ for  SBH-only simulations (where $N_\mathrm{0,NS}=0$ is assumed), and $M_0=0.1M_\odot$ for simulations that include both BHs and NSs. Our results are not sensitive to these  choices of $M_0$.

Finally, we should note that natal kicks of the compact objects can significantly impact not only  the normalization of the IMF (i.e. $N_\mathrm{0,SBH}+N_\mathrm{0,NS}$), but also  its shape, as these kicks are mass dependent and mainly affect the lighter objects. Yet, in the name of simplicity and tractability of the results, we choose to work with a relatively simple model where the normalization of the IMF is governed by the free parameters $N_\mathrm{0,SBH}$ and $N_\mathrm{0,SBH}$, while its shape is determined solely by the power-law $\alpha$.

\subsection{Interaction Rate}
As mentioned above, the rate $R_{i,j}$ consists of two components, the interaction kernel $\tilde{R}_{i,j}$ and the overall interaction rate regulator $R_{11}$. We treat $R_{11}$ as a free parameter in our model. 

Ideally, we would strive to use an analytic formula for the interaction rate kernel derived from first principles.
An essential property that every physical kernel $\tilde{R}_{i,j}$ must hold is symmetry with respect to the indices $i$ and $j$, otherwise the corresponding cross section will not be symmetrical to mass exchange which is forbidden due to Lorentz invariance. The only 2-argument fundamental functions which are symmetric under the exchange of their arguments are addition and multiplication. Physical cross sections are characterized by power-law functions, thus the simplest physical form for the kernel is $\tilde{R}_{i,j}\propto\left(i\cdot j\right)^a\left(i+j\right)^b$ --- indeed this is the same parameterization that was used in \cite{Christian:2018mjv}.

It was noted in \cite{Mouri:2002mc, Sigurdsson:1994ju} that mass segregation (a process by which heavier objects sink towards the cluster core due to dynamical friction) gives the index $a$ an additional contribution of 3/2, assuming a King density profile. In appendix \ref{Derivation for 2-body capture rate kernel} we show that $a=15/14$, $b=9/14$ correspond to two-body gravitational capture processes (without mass segregation). However, such kernel ignores 3-body encounters, which likely dominate the merger rate in GC environments~\citep{Lee:1993}. As a chaotic system, there is no closed formula for the 3-body interactions kernel, although there were attempts to fit numerical simulations ~\citep{Christian:2018mjv, OLeary:2016ayz}. Therefore, we choose to characterize the kernel using a phenomenological model, given by \cite{Kovetz:2018vly, Doctor:2019ruh, Fishbach:2019bbm, Fishbach:2020ryj}
\begin{equation}\label{pheno kernel}
\tilde{R}_{i,j}=\left[\mathrm{min}\left(\frac{i}{j},\frac{j}{i}\right)\right]^\beta\left(\frac{i+j}{2}\right)^\gamma.
\end{equation}
The indices $\beta$, $\gamma$ which appear in Eq.~(\ref{pheno kernel}) are free parameters. The more positive $\beta$ is, the more the kernel prefers interactions with equal masses. For negative $\beta$, the kernel prefers asymmetric-mass interactions. When $\gamma$ is greater, there are more interactions that involve heavier BHs. When $\beta=\gamma=0$, the kernel is simply unity, which means that it has no preference for certain interactions over others. In \cite{Doctor:2019ruh} it was estimated that $\gamma\sim2$, as the rate is expected to be proportional to $\left(r_i+r_j\right)^2$, where $r_i$, $r_j$ are the Schwarzschild radii of the BHs. Realistically, mass segregation in GCs naturally predicts a positive $\beta$ \citep{Sollima:2008ei, Park:2017zgj, Rodriguez:2019huv}. Since the heavier BHs are concentrated within a smaller volume in the GC core, $\gamma$ would also tend to be positive.

\subsection{Delay time}
For a given binary, the delay time $t_d$ between the formation time and the merger time depends on various properties, including metallicity \citep{Mandel:2015qlu, Marchant:2016wow, Buisson:2020hoq}, the nature of the interaction \citep{Antonini:2017ash, Rodriguez:2018jqu, Hoang:2017fvh}, and the environment \citep{Banerjee:2009hs, Benacquista:2011kv, Rodriguez:2016kxx, Banerjee:2016ths, Rodriguez:2017pec, Samsing:2017xmd, Kremer:2019iul, DiCarlo:2020lfa}. In this work we shall adopt the conventional delay time distribution of $t_d^{-1}$ with a minimum delay time $t_{d,\mathrm{min}}$,
\begin{equation}\label{delay time distribution}
P\left(t_d;t\right)\propto\frac{1}{t_d}\mathcal{H}\left(t-t_d\right)\mathcal{H}\left(t_d-t_{d,\mathrm{min}}\right).
\end{equation}
As was noted by \citet{Fishbach:2021mhp}, there could be some dependence of $t_{d,\mathrm{min}}$ on the masses of the binary. In most of our simulations we shall adopt a global $t_{d,\mathrm{min}}=50\,\mathrm{Myears}$ \citep{Kovetz:2018vly} for all the binaries, except for appendix \ref{Mass dependent td,min} where we discuss the implications of mass dependent $t_{d,\mathrm{min}}$.

Naturally, the distribution of Eq.~\eqref{delay time distribution} has to be properly normalized. Sadly, normalizing this distribution at $t\to\infty$ is impossible because the integral $\int_{t_{d,\mathrm{min}}}^\infty t_d^{-1}dt_d$ does not converge, thus one must impose a maximal delay time. We have chosen to work with $t_{d,\mathrm{max}}=2t_c=26\,\mathrm{Gyears}$ and we have verified that our results are not sensitive to this particular choice.

\subsection{Mass loss}
For equal-mass BHs, the merger product is $\sim95\%$ of their sum~\citep{HolleyBockelmann:2007eh, Pretorius:2005gq, Campanelli:2005dd, Baker:2006yw, Herrmann:2007ac}, which means that $\sim5\%$ of the initial energy is channeled to GWs. The energy loss however is smaller when the BHs are not equal in size. We follow \cite{Flanagan:1997sx} and model the energy carried away by GWs as
\begin{equation}\label{E_GW}
E_\mathrm{GW}\left(M_i,M_j\right)=16\cdot f_\mathrm{loss}\frac{\left(M_i\cdot M_j\right)^2}{\left(M_i+M_j\right)^3},
\end{equation}
and round $E_\mathrm{GW}$ to the nearest multiple integer of $M_0$. The parameter $f_\mathrm{loss}$ in Eq.~\eqref{E_GW} is the fractional energy loss when $M_i=M_j$. For large mass asymmetry, $E_\mathrm{GW}\left(M_i,M_j\right)$ is very small, reflecting the fact that only a small fraction of the lighter BH mass is transformed into GWs. 
For SBHs mergers the value of $f_\mathrm{loss}$ is expected to be 0.05. 
We note that this small value does not have a great impact on the events distribution and therefore we set it to be zero at section \ref{Comparison to observations} --- but we do show its impact when we study the growth in the cluster at section \ref{Coagulation results}. For NSs $f_\mathrm{loss}$ is expected to be very small as the energy loss in binary NS is negligible~\citep{Gupta:2019nwj, Shibata:2019wef}.

\subsection{Ejections}

In appendix \ref{Derivation for the relation between the probability for ejection and the recoil velocity} we derive the relation between the probability for ejection $P_\mathrm{ej}$ and the recoil velocity $v_\mathrm{rec}$,
\begin{equation}\label{Probability for ejection}
P_{\mathrm{ej}}=\begin{cases}
\displaystyle{1-\left(1-\frac{v_\mathrm{rec}^4}{v_\mathrm{esc}^4}\right)^{3/2}} & v_\mathrm{rec}\leq v_\mathrm{esc}\\
1 & v_\mathrm{rec}\geq v_\mathrm{esc}
\end{cases}.
\end{equation}

As mentioned previously, BHs might gain a large recoil velocity during their merger as a result of asymmetry in the radiated GWs. This asymmetry (GWA) can be due to the spin orientations of the BHs or due to different masses of the coalescing BHs. Neglecting the spins contribution, the expression for the recoil velocity as the result of mass asymmetry is $v_\mathrm{rec}\propto\frac{q^2\left(1-q\right)}{\left(1+q\right)^5}\left[1-0.93\frac{q}{\left(1+q\right)^2}\right]$ where $q$ is the mass ratio. 
The complete expression that accounts for the BHs spins as well can be found in \cite{HolleyBockelmann:2007eh, Fragione:2017blf}. As we do not track in this work the BHs spins during the evolution of the cluster, we set all the binaries to have the same spin values (magnitude and orientation).

In the case of ejections due to 3-body interactions, the recoil velocity of the center of mass is given in \cite{Antonini:2018auk, Miller:2008yw}. We use a simpler model which shares the same mass dependence in the limits of very light or heavy mass\footnote{Namely, $v_\mathrm{rec}\propto M_i$ for $M_i\ll M_j$, and $v_\mathrm{rec}\propto M_i^{-1}$ for $M_i\gg M_j$.},
\begin{equation}\label{recoil velocity 3-body}
\frac{v_\mathrm{rec}^2\left(M_i,M_j\right)}{v_\mathrm{esc}^2}=\frac{M_iM_j^2}{M_\mathrm{esc}\left(M_i+M_j\right)^2},
\end{equation}
where $M_\mathrm{esc}\equiv av_\mathrm{esc}^2/0.2G$ and $a$ is the initial semi-major axis of the binary. $M_\mathrm{esc}$ can be thought of in the following way: for a large mass ratio ($M_i\ll M_j$), the lighter object will always get ejected if $M_i>M_\mathrm{esc}$. 

In our model, $M_\mathrm{esc}$ and $v_\mathrm{esc}$ serve as free parameters that control the amount of ejections for 3-body and GWA types, respectively\footnote{For an ejections-free GC, we set $M_\mathrm{esc}\!=\!v_{\mathrm{esc}}\!=\!\infty$,  yielding $P_{\mathrm{ej}}\!=\!0$.}. In order to understand the effect of each of the two parameters, in all of our simulations we consider only one type of ejections at a time --- either 3-body or GWA, but not both.

\section{Coagulation results}\label{Coagulation results}

\begin{figure}
\includegraphics[width=\columnwidth]{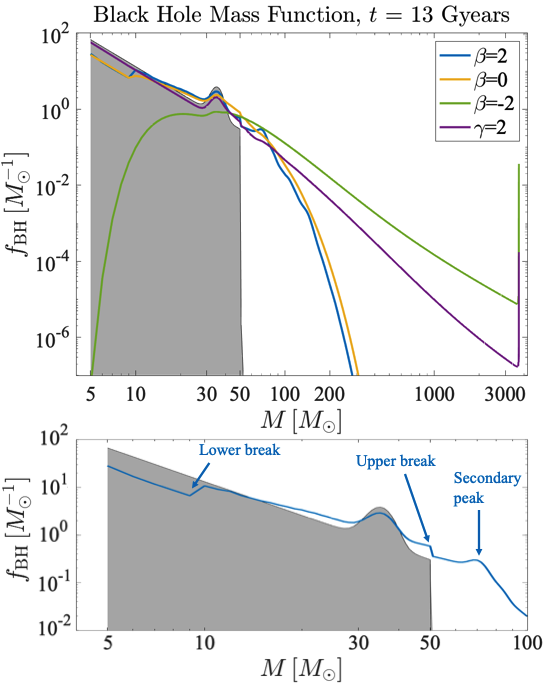}
\caption{Upper panel: The cluster BHMF after $13\,{\rm Gyrs}$ of coagulation for different $\beta$ and $\gamma$ values, with $N_\mathrm{0,SBH}=300$, no NSs, and no ejections nor mass loss (i.e., $N_\mathrm{0,NS}=0$, $v_\mathrm{esc}=M_\mathrm{esc}=\infty$, and $f_\mathrm{loss}=0$) and zero delay time. For each curve, the value of $R_{11}$ was chosen to match the total number of events in  O3a (see section \ref{Comparison to observations}). The black curve corresponds to the IMF (Eq.~\eqref{stellar IMF}). In each scenario we consider either $\beta\neq0$ or $\gamma\neq0$, but not both. Lower panel: zooming-in on the $\beta=2$ curve.}
\label{figure_2}
\end{figure}

We start this section by focusing only on the BH population in the cluster, whose IMF is given in Eq.~\eqref{stellar IMF}. The most critical quantities that affect the cluster evolution are the kernel power-law indices $\beta$ and $\gamma$. In Fig.~\ref{figure_2} we plot the final BHMF for different choices of these parameters, while assuming no mass loss nor ejections and zero delay time. We can see that the $\beta=2$ curve exhibits some interesting properties. These are the lower break at $\sim10M_\odot$, the upper break at $\sim50M_\odot$ and the secondary peak at $\sim70M_\odot$. The lower break is the result of the IMF low-mass cutoff. Since there are no BHs below $5M_\odot$, only BHs above $10M_\odot$ can be formed via mergers, which causes a sudden rise at $10M_\odot$. The secondary peak is the result of mergers of BHs at the PPSN peak with themselves, resulting in remnant BHs which are twice as heavy. These two properties are most manifest for $\beta=2$ as this kernel strongly favors equal-mass mergers. The upper break is the result of the IMF high-mass cutoff and the moderate expansion of the BHMF to high masses for high $\beta$ values.

When $\beta$ becomes lower (and more negative), the rapid growth smooths the discontinuous initial upper cutoff. When $\beta\!=\!-2$, the kernel favors merging very heavy masses with very light ones, thus causing  runaway growth in the heavy-mass domain on the one hand and runaway \emph{depletion} in the light-mass domain on the other hand. $\gamma$ also has significant impact on the coagulation process. For $\gamma>0$ there is a strong preference for heavy-mass mergers, resulting in extremely rapid runaway growth.

\begin{figure}
\includegraphics[width=\columnwidth]{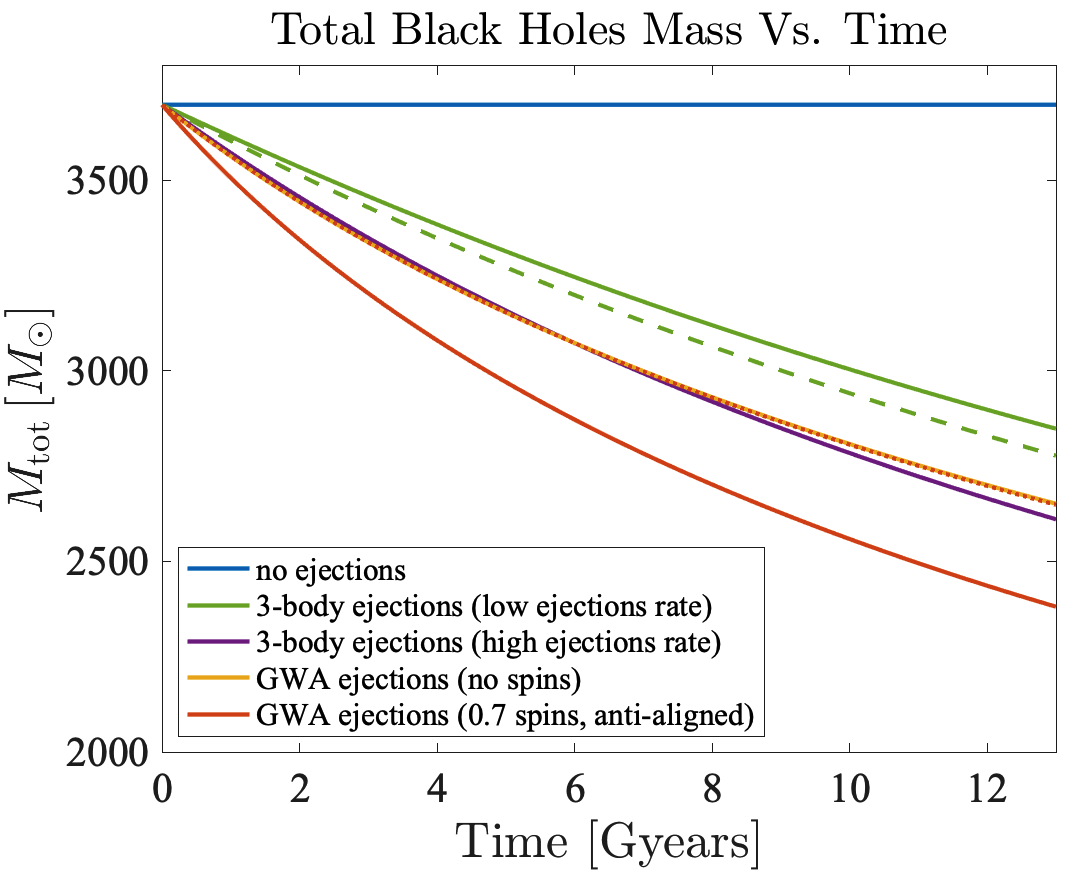}
\includegraphics[width=\columnwidth]{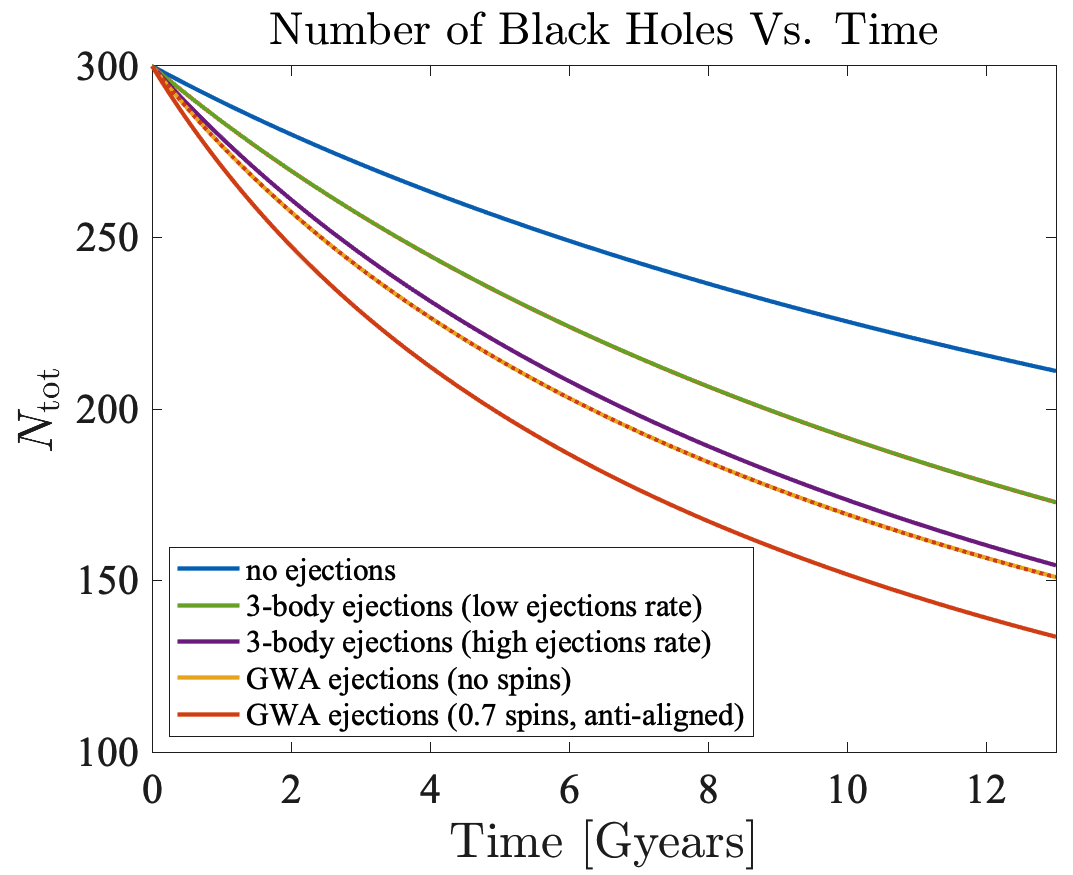}
\caption{Top (bottom) panel: $M_\mathrm{tot}$ ($N_\mathrm{tot}$) as function of time. We set $M_\mathrm{esc}=5M_\odot$ ($M_\mathrm{esc}=3M_\odot$) for the low (high) 3-body ejections rate scenario. For the GWA scenarios we set $v_\mathrm{esc}=50~\mathrm{km/s}$. All solid curves correspond to $f_\mathrm{loss}=0$, while the green dashed line corresponds to $f_\mathrm{loss}=0.05$. Here we used $\beta=2$, $\gamma=0$, $N_\mathrm{0,SBH}=300$, $N_\mathrm{0,NS}=0$ and zero delay time. The value of $R_{11}$ is the same for all curves and  was chosen to match the O3a total number of events (see section \ref{Comparison to observations}) with respect to the \emph{blue} curves (no ejections nor mass loss).  The red dotted curve corresponds to a "mixed spins" scenario where BHs below $50M_\odot$ are spin-less and BHs above $50M_\odot$ have spins (both in magnitude and in orientation) as in the solid red curve (this curve ovelaps the yellow "no spins" curve).}
\label{figure_3}
\end{figure}

We can also learn about the GC evolution by monitoring how its statistical properties change over time. For example, we can track the total number $N_\mathrm{tot}$ and mass $M_\mathrm{tot}$ in BHs that remain in the GC. 
This is done in Fig.~\ref{figure_3}. As expected, $M_\mathrm{tot}$ is a conserved quantity when there are no ejections nor mass loss, but $N_\mathrm{tot}$ decreases as BHs merge. 
When ejections are considered, both $M_\mathrm{tot}$ and $N_\mathrm{tot}$ can decrease over time. When $f_\mathrm{loss}\neq0$ is also included (green dashed curve), we see a minor decrease in $M_\mathrm{tot}$ but $N_\mathrm{tot}$ remains the same (compared to the case where $f_\mathrm{loss}=0$), as expected.

 For 3-body ejections, we see that the amount of ejections is increased when $M_\mathrm{esc}$ is decreased (in this example we set the value of $M_\mathrm{esc}$ to be either $3M_\odot$ or $5M_\odot$). For the GWA ejections we consider 50~km/s, which is a typical escape velocity for GCs ~\citep{Merritt:2004xa, Favata:2004wz, HolleyBockelmann:2007eh}.  For an escape velocity of $\sim100$~km/s, the GWA ejection rate coincides with the low 3-body ejections rate (in \cite{Kimball:2020qyd}, mergers in clusters were  found to be consistent with the LIGO O3a results if the escape velocity is indeed $\gtrsim100$~km/s). Besides the mass ratio asymmetry contribution to the ejections (yellow curve), the BH spins increase the ejection rate even further (red curve) \citep{Rodriguez:2019huv}. As an extreme scenario, we chose to examine the scenario where all of the BHs dimensionless spin is 0.7 and spin configuration is such that it yields the most ejections (when the spin of one BH is aligned with the orbital angular momentum of the binary and the other BH spin is anti-aligned to it). We also examine a more mundane scenario where we assume that all the BHs below $50M_\odot$ (our upper mass cutoff) are spin-less and all the BHs above $50M_\odot$ have 0.7 dimensionless spins. This is of course not an accurate description of the spin distribution in the cluster as second generation BHs which are below $50M_\odot$ might gain a high spin value, but this choice allows us to utilize our prescription without modification to monitor the BHs spins as they merge. This "mixed spins" scenario is depicted in Fig.~\ref{figure_3} by a dotted red curve. This curve overlaps the yellow curve of the "no-spins" scenario, thus we learn that the majority of the second generation BHs are indeed ejected. As we shall demonstrate in section \ref{Comparison to observations}, the few retained second generation BHs are detectable.

Our prescription also allows us to calculate the probability to form an IMBH throughout the GC lifetime.  In order to do so, we first need to consider the probability that a randomly selected BH in our cluster is an IMBH. This probability is given by
\begin{equation}\label{P_500}
P_{500}=\nicefrac{\displaystyle{\sum_{i=i_{500}}^{i_\mathrm{max}}N_i}}{\displaystyle{\sum_{i=1}^{i_\mathrm{max}}N_i}},
\end{equation}
where $i_{500}=500M_\odot/M_0$. 
To estimate the probability that an IMBH has formed in the cluster, we ``check" if at least one of the $N_\mathrm{tot}$ BHs is an IMBH, which gives
\begin{equation}\label{P_IMBH}
P_\mathrm{IMBH}=1-(1-P_{500})^{N_\mathrm{tot}}.
\end{equation}

\begin{figure}
\includegraphics[width=\columnwidth]{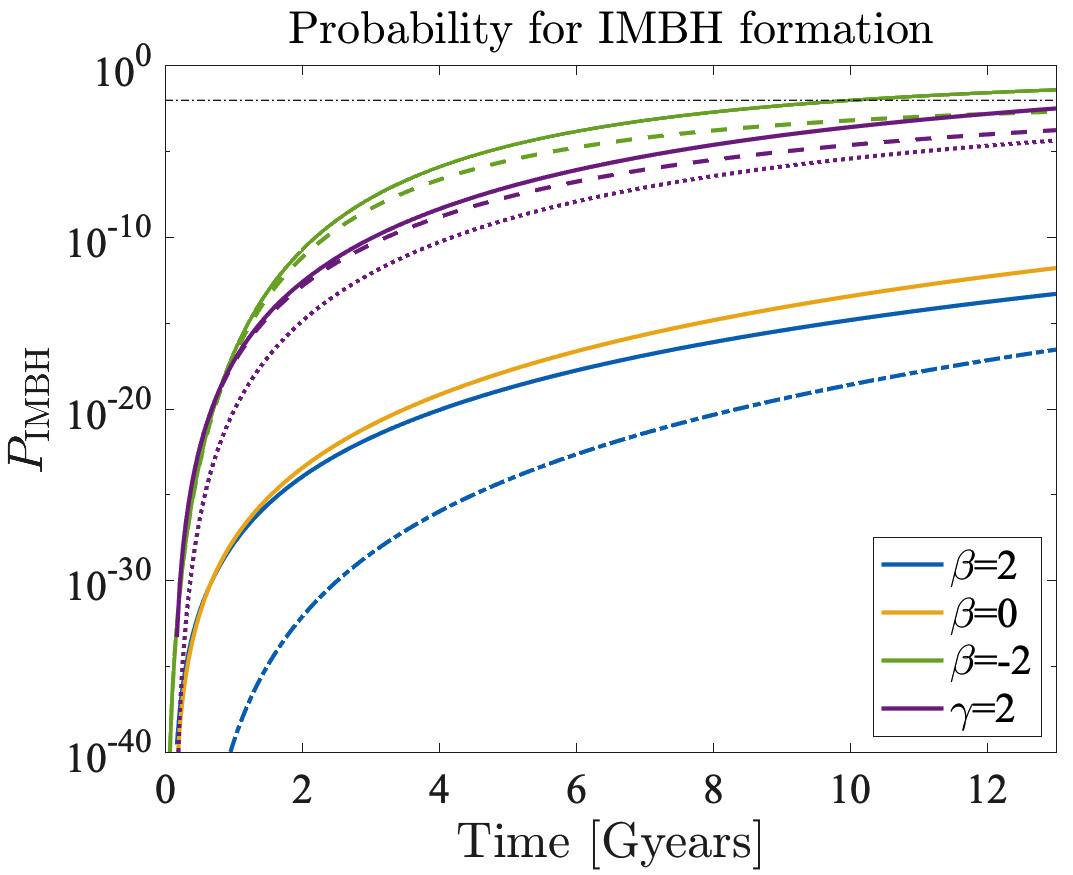}
\caption{Probability to form an IMBH for different $\beta$ and $\gamma$ values, with $N_\mathrm{0,SBH}=300$, $N_\mathrm{0,NS}=0$. In each scenario we consider either $\beta\neq0$ or $\gamma\neq0$. \emph{Solid} lines correspond to no ejections nor mass loss ($v_\mathrm{esc}=M_\mathrm{esc}=\infty$, $f_\mathrm{loss}=0$) and zero delay time. \emph{Dashed} lines correspond to 3-body ejections but no mass loss ($M_\mathrm{esc}=5M_\odot$, $f_\mathrm{loss}=0$) and no delay time. The purple \emph{dotted} line corresponds to both ejections and mass loss ($M_\mathrm{esc}=5M_\odot$, $f_\mathrm{loss}=0.05$), but still without any delay time. The blue \emph{dashed-dotted} curve corresponds to no ejections nor mass loss but the delay time is now considered. For each \emph{solid} curve, the value of $R_{11}$ was chosen to match the O3a number of events (see section \ref{Comparison to observations}), and its value remains the same when introducing either ejections or mass loss. The black \emph{dot-dashed} line represents the Milky Way (MW) threshold value: assuming a total of $\sim100$ MW GCs, it is likely that an IMBH exists in the MW when $P_\mathrm{IMBH}$ surpasses $10^{-2}$.}
\label{figure_4}
\end{figure}

\noindent Fig.~\ref{figure_4} presents $P_\mathrm{IMBH}\left(t\right)$ for the same $\beta$ and $\gamma$ values we used in Fig.~\ref{figure_2}. As more and more BHs merge to form heavy BHs, $P_\mathrm{IMBH}$ becomes greater as time progresses. For $\beta=0,2$ the probability is too low to consider  IMBH formation via coalescence in GCs. For $\beta=-2$ the probability grows much faster and reaches order of ten percent by the end of the coagulation period. For $\gamma=2$ the probability grows much faster (compared to $\gamma=0$) due to the extreme runaway growth. However, when low rate ejections are included (here we worked with the same $M_\mathrm{esc}$ value of the low ejections rate curve in Fig.~\ref{figure_3}), the probability drops down by an order of magnitude\footnote{If we had used a smaller value for $M_\mathrm{esc}$, the probability would drop down more significantly and may even vanish at some point as not enough mass remains in the cluster to form an IMBH.}. When mass loss is also considered, the probability decreases by an additional order of magnitude. Finally, we also present in Fig.~\ref{figure_4} how the delay time affects the IMBH formation (for $\beta=2$). Clearly, its formation is delayed and $P_\mathrm{IMBH}$ consistently decreases even further by several orders of magnitude.

\begin{figure}
\includegraphics[width=\columnwidth]{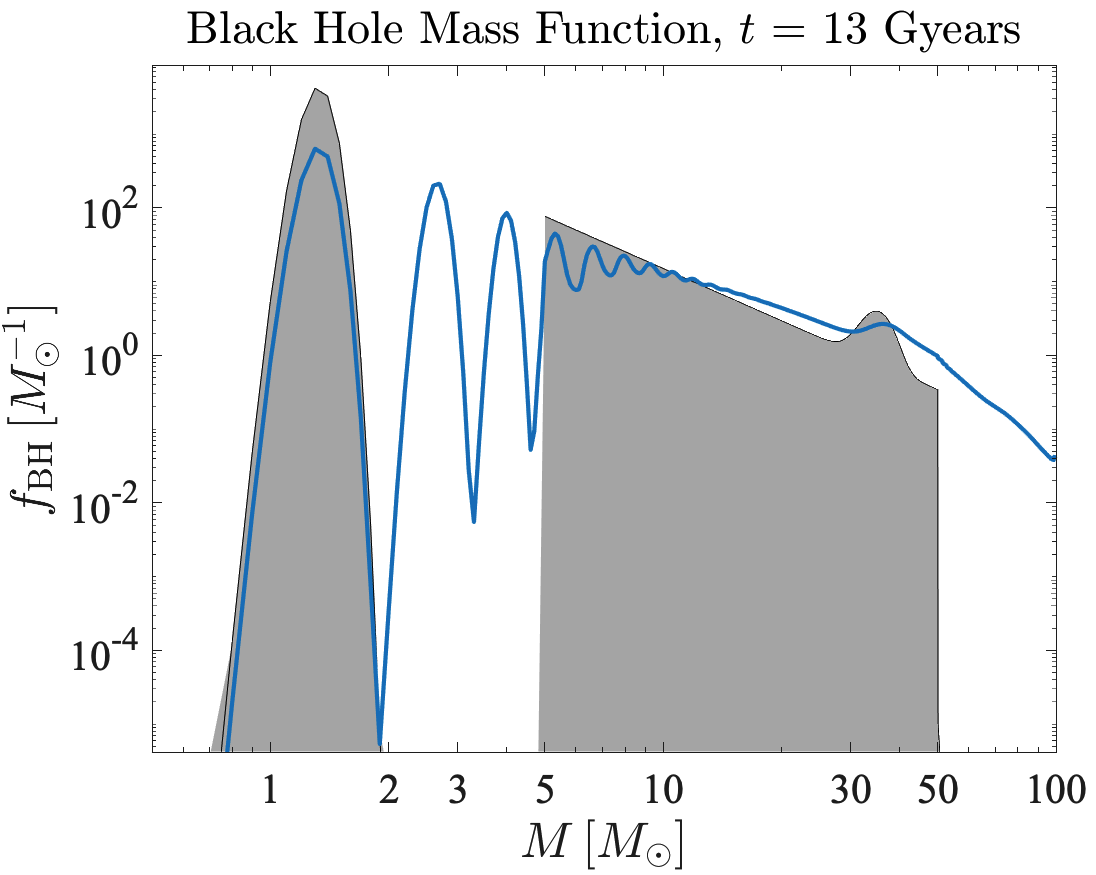}
\caption{\emph{Blue curve}: the GC BHMF for $\beta=\gamma=0$, $N_\mathrm{0,SBH}=300$, $N_\mathrm{0,NS}=1000$, no mass loss nor ejections and zero delay time. \emph{Black curve}: the combined IMF of SBHs and NSs (Eq.~\eqref{SBH+NS IMF}).}
\label{figure_5}
\end{figure}

We now add the NS contribution to the IMF and solve the coagulation equation for $\beta=\gamma=0$. The result is shown in Fig.~\ref{figure_5}. Within the lower mass gap, the wiggly shape of the curve is due to the narrow NS IMF. Most of the NSs are concentrated around $1.33\,M_\odot$, and as there is no mass loss in NS collisions, most of the 2nd-generation BHs are clustered around the second peak located at $2.66\,M_\odot$ and in a similar manner the distance between all adjacent peaks is roughly $\sim1.33\,M_\odot$. The wiggles decay in amplitude and gradually smooth out as they enter the SBH regime above $5\,M_\odot$.

\begin{figure}
\includegraphics[width=\columnwidth]{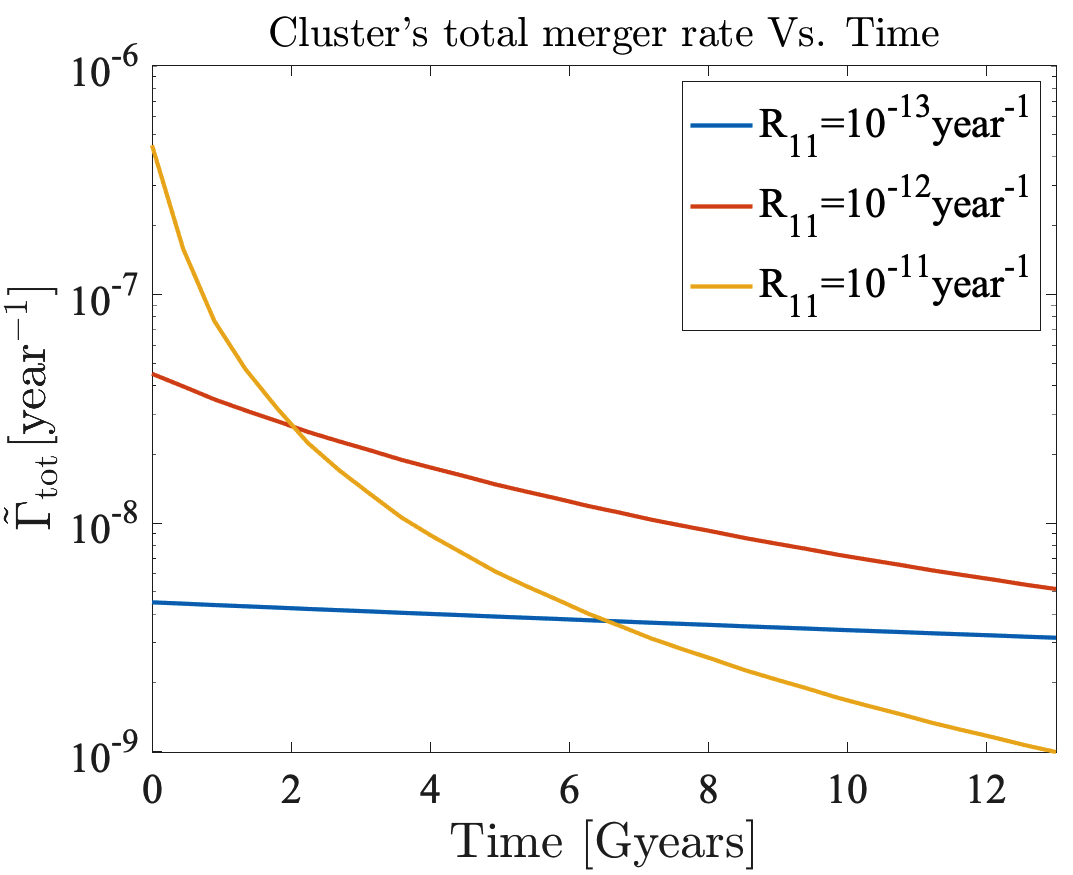}
\caption{The overall GC merger rate for different $R_{11}$ values; $\beta=\gamma=0$, $N_\mathrm{0,SBH}=300$, $N_\mathrm{0,NS}=0$, no mass loss nor ejections and zero delay time.}
\label{figure_6}
\end{figure}

A useful quantity to estimate the total number of detected events is the overall merger rate, defined as
\begin{equation}\label{Gamma_tot}
\tilde\Gamma_\mathrm{tot}\left(t\right)=\frac{1}{2}\sum_{j=1}^{i_{\mathrm{max}}}\sum_{k=1}^{i_{\mathrm{max}}}\tilde\Gamma_{j,k}\left(t\right)\Theta_{i_\mathrm{max},i_\mathrm{rem}},
\end{equation}
where $\Theta_{i_\mathrm{max},i_\mathrm{rem}}$ is the discrete step function and its appearance is required in order to exclude mergers with remnants that are heavier than the GC mass in BHs. Fig.~\ref{figure_6} presents $\tilde\Gamma_\mathrm{tot}\left(t\right)$ for different $R_{11}$ values. As expected, greater $R_{11}$ results in a greater initial overall merger rate. However, when $R_{11}$ is too large, it can lower the final overall merger rate, because most of the BHs have already merged at the beginning and the cluster ends up with only a few BHs that can still merge.

\begin{figure}
\includegraphics[width=\columnwidth]{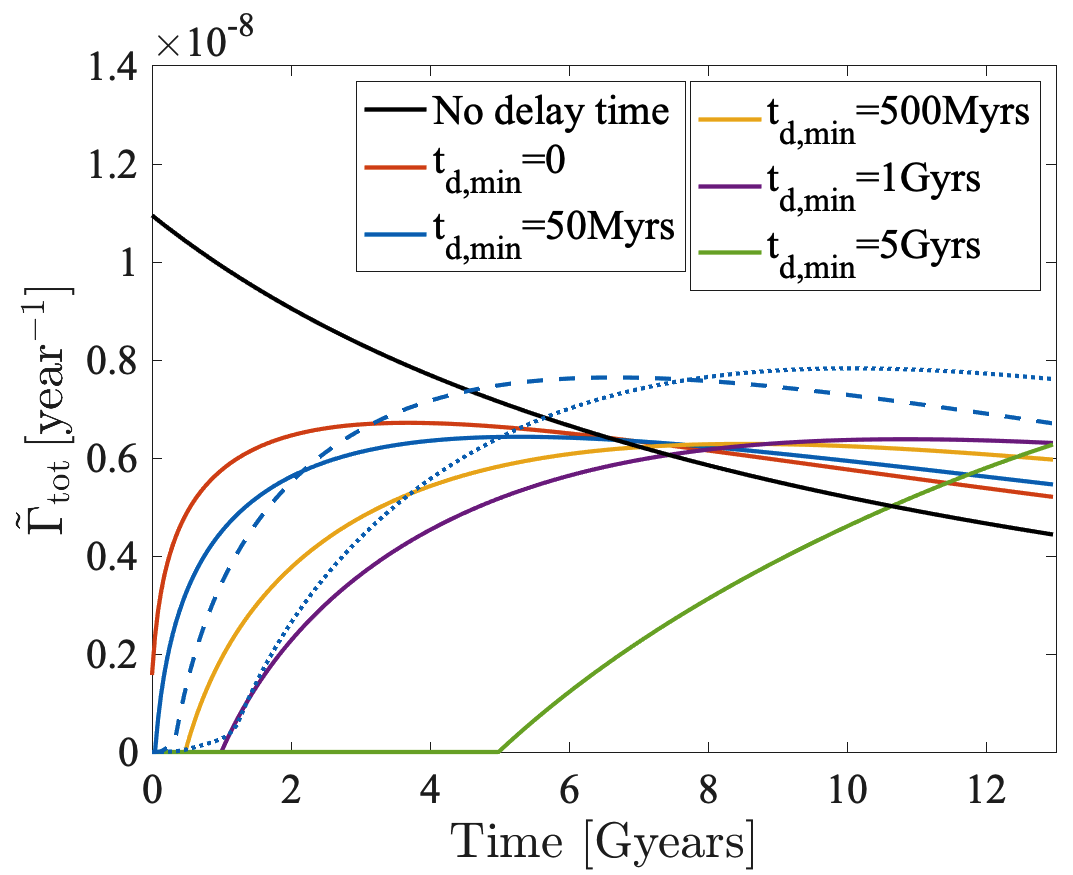}
\caption{The overall GC merger rate for different $t_{d,\mathrm{min}}$ values; $\beta=2$, $\gamma=0$, $N_\mathrm{0,SBH}=300$, $N_\mathrm{0,NS}=0$, no mass loss nor ejections. The dashed (dotted) curve corresponds to a mass dependent $t_{d,\mathrm{min}}$ with a slope of $7.14\,\mathrm{Myears}/M_\odot$ ($27.14\,\mathrm{Myears}/M_\odot$) --- see details and discussion at appendix \ref{Mass dependent td,min}. The value of $R_{11}$ is the same for all the curve and was chosen to match the O3a number of events in the no delay time scenario.}
\label{figure_7}
\end{figure}

Lastly, we explore in Fig.~\ref{figure_7} the impact of the delay time on the total merger rate. The black curve corresponds to the no delay time scenario. Here, the BHs are allowed to merge at $t=0$, and the merger rate decays with time as the number of BHs decreases. When the delay time is introduced, the initial merger rate is zero, and then gradually increases as more and more binaries finish their "waiting" period. The merger rate reaches a maximal value once the cluster has lost enough BHs. As expected, the derivative of $\tilde\Gamma_\mathrm{tot}$ is smaller for greater values of $t_{d,\mathrm{min}}$. Interestingly, for the values of $t_c$ and $R_{11}$ that we have chosen to work with, the final merger rate has little dependence on $t_{d,\mathrm{min}}$.

\section{The observed merger density function}\label{The observed merger density function}
In the previous sections we explained that one can obtain the current GC BHMF $N_i\left(t_\mathrm{c}\right)$ from the IMF $N_i\left(t=0\right)$ by solving the coagulation equation. $N_i\left(t_\mathrm{c}\right)$ however corresponds only to the BHMF  at the source frame. It is different than the observed BHMF which is inferred from aLIGO measurements. The observed BHMF is biased due to several effects, described below.

First and most importantly, we are not sensitive to the BHs themselves but rather to their \emph{mergers}. This means that even if the true BHMF $N_i\left(t\right)$ is high at some mass values, if these BHs do not merge often we will not be able to detect them. Secondly, the strain signal which is measured by aLIGO is roughly proportional to the reduced mass of the two BHs~\citep{Flanagan:1997sx}. Since the signal has to overcome the noise background in order to be detected, this effect implies that aLIGO is more sensitive to heavier BHs. Thirdly, the detected strain signal is inversely proportional to the distance of the merger from Earth and thus farther events are less detectable. Fourthly, the strain frequencies are highly dependent on the masses involved in the merger. Since the aLIGO noise is not constant in frequency, this effect implies that different merging masses yield  different signal-to-noise ratios. Finally, measurement errors might affect the deduced masses from the strain signal. All of these effects have to be considered when predicting the observed BHMF.

During its runs, aLIGO can detect BH merger events from different origins around the cosmos. We consider two types of origins, or \emph{channels}. The first type of mergers come from \emph{dynamic} environments, e.g.\ GCs. In such environments the merger rate changes over time due to changes in the population. These changes can be modelled via the coagulation equation as we have discussed. The second type of environments that we consider are \emph{static}, i.e.\ mergers occurring in the galactic field. These are static in the sense that their mass function does not change considerably over time due to interactions\footnote{ However, the merger rate of the static channel may be time-dependent and evolve with redshift - see details underneath Eq.~\eqref{f_det(M1,M2,z)^stat}.}. Below we present our prescription to analyze the contributions of each channel to the detected merger events.

\subsection{Dynamic channel}
Let us focus first on the merger events that take place in the dynamic channel, namely inside GCs (or other dense clusters). In principle, the properties of the GCs that host these mergers may vary --- such as their age, density profile, the number of compact objects they host and their velocity dispersion, etc.
Observations suggest that most GCs are old~\citep{Valcin:2020vav}, a statement which is also supported by theoretical considerations (see appendix \ref{Derivation for the GC age distribution}), allowing us to assume that all GCs are $\sim13\,{\rm Gyrs}$ old. As for the other GC properties, we assume that our simulated GC  represents the typical GC whose mergers are observed by aLIGO. It is straightforward to extend our methodology to entertain different populations of GCs, a task we leave to future work.

The most important quantity that aLIGO is sensitive to during the GC lifetime is the merger rate $\tilde\Gamma_{i,j}\left(t\right)$, which is obtained by performing a convolution between $P\left(t_d\right)$ and the interaction rate (given by Eq.~\eqref {Gamma_i,j}). It is straightforward to convert this discrete quantity into a continuous rate-density function $\tilde\Gamma\left(M_1,M_2,t\right)$ (with $M_1\geq M_2$). We convert time $t$ to redshift $z$ using standard $\Lambda$CDM cosmology~\citep{Aghanim:2018eyx}, yielding $\tilde\Gamma\left(M_1,M_2,z\right)$.

\begin{figure*}
\includegraphics[width=\columnwidth]{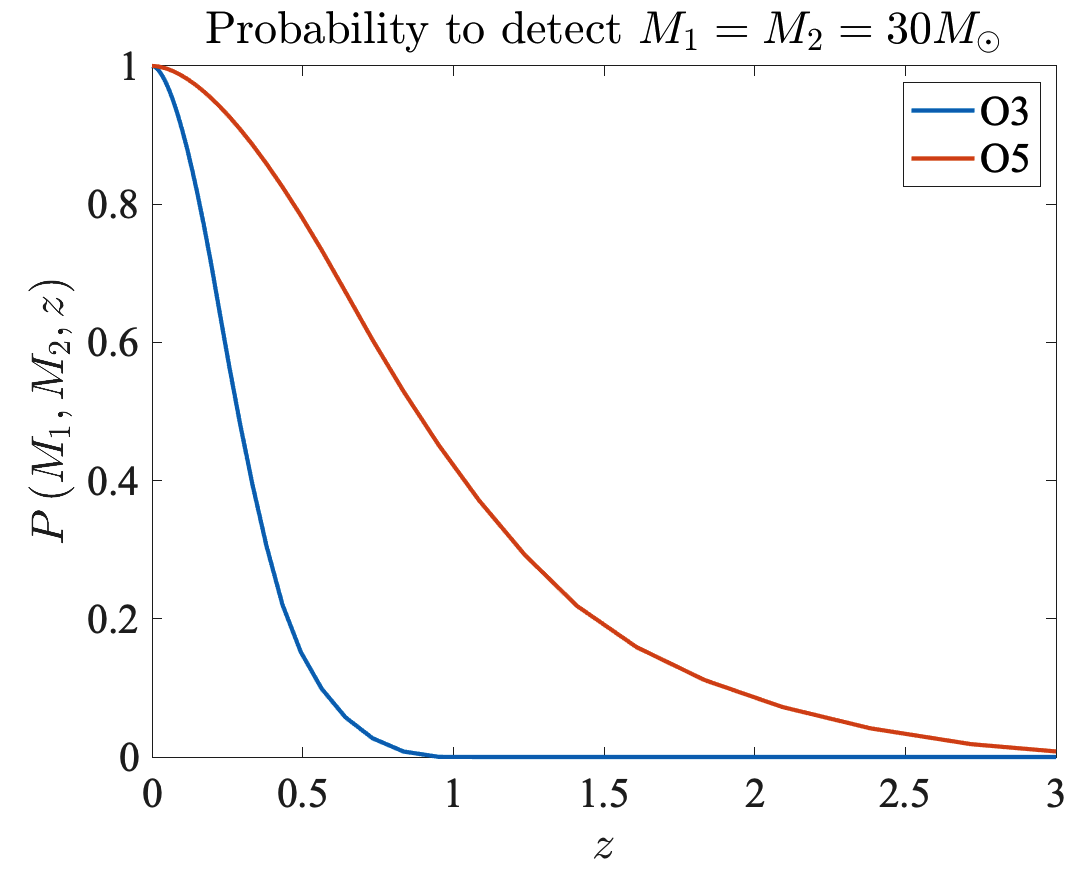}
\includegraphics[width=\columnwidth]{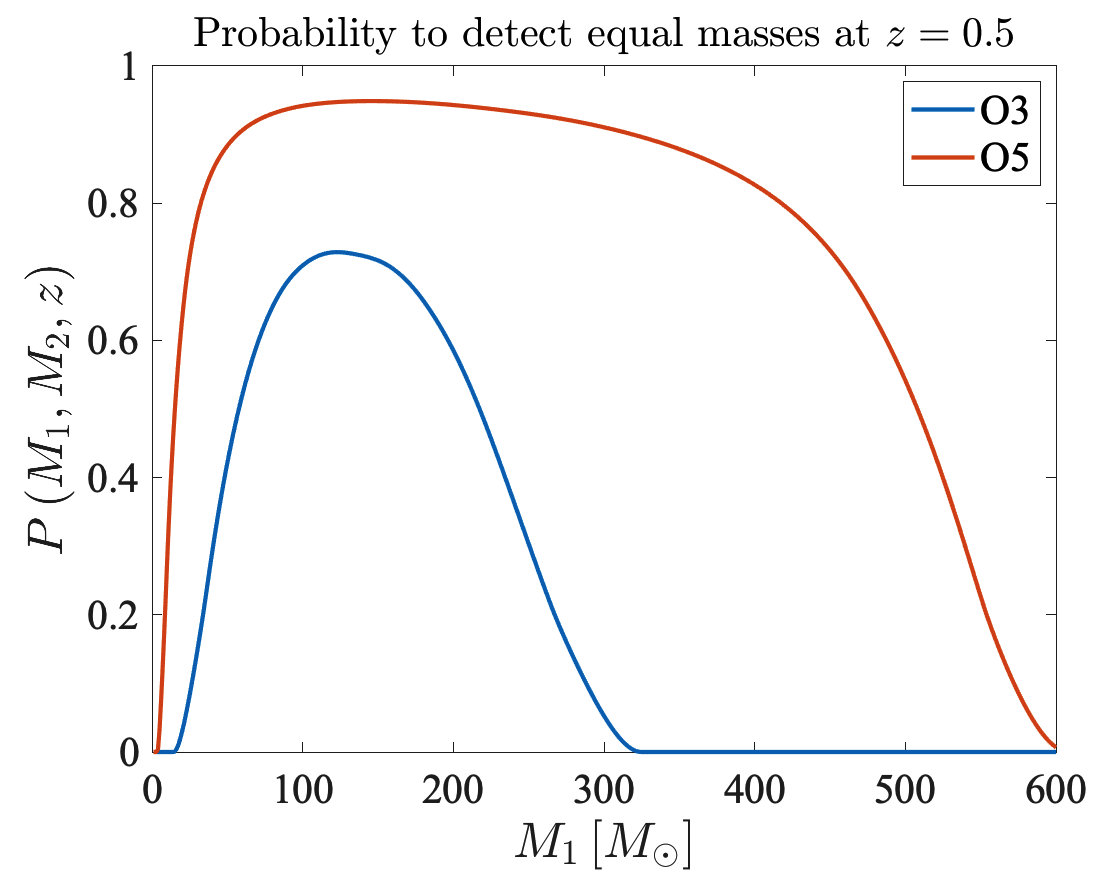}
\caption{Left panel: Probability to detect $M_1=M_2=30M_\odot$ mergers as a function of the events redshift $z$. Right panel: Probability to detect equal masses mergers at redshift $z=0.5$ as a function of the masses.}
\label{figure_8}
\end{figure*}

Consider the number of mergers with mass components in the ranges $\left[M_1,M_1+\mathrm{d}M_1\right]$ and $\left[M_2,M_2+\mathrm{d}M_2\right]$ that occur within a single GC located at redshift $z$ during a total observation time period $T$. This number is given by $T\times\tilde\Gamma\left(M_1,M_2,z\right)\mathrm{d}M_1\mathrm{d}M_2$. However, this number reflects only the number of mergers as seen at source frame (near the cluster). The observed merger rate is different due to two effects. First, since the universe expands, the observed rate is the rate at source frame times $1/\left(1+z\right)$ \citep{Fishbach:2018edt}. Secondly, the signal-to-noise $\rho$ has to surpass a certain threshold to be detected (we use a threshold of $\rho=8$ per detector for O3a and a network threshold of $12/\sqrt{4}$ ~\citep{Thorne:1987af, Allen:2005fk, Chen:2017wpg} for O5 where 4 detectors were assumed). However, even when this threshold is surpassed there is still non-zero probability that the signal will not be detected, due to the configuration of the detectors. The detection probability is a function of  $\rho$ alone\footnote{The signal-to-noise is also a function of the inclination and orientation angles as well as the BH spins. However, since mergers are not likely to have a preferred direction, the mean value for all these quantities is zero.}, which depends on the masses $M_{1},M_{2}$ and the event redshift $z$, so we denote this probability as $P\left(M_{1},M_{2},z\right)$\footnote{This probability also accounts for the redshifting in the GW frequencies due to the expansion of the universe.}. Therefore, the number of detected mergers from a single GC at redshift $z$ is given by
\begin{eqnarray}\label{dN per GC at detector}
\nonumber\mathrm{d}N_{\mathrm{mergers/GC}}^{(\mathrm{det})}=&&\frac{1}{1+z}T\times\tilde\Gamma\left(M_1,M_2,z\right)
\\&&\times P\left(M_{1},M_{2},z\right)\mathrm{d}M_1\mathrm{d}M_2.
\end{eqnarray}

In order to find the number of events that aLIGO will detect, we ought to estimate the cosmological abundance of GCs as a function of $z$. 
First, consider the comoving volume element
$\mathrm{d}V_\mathrm{c}=\chi^2\left(z\right)\mathrm{d}\chi\mathrm{d}\Omega$,
where $\chi\left(z\right)$ is the comoving distance to redshift $z$. 
The number of GCs at redshift $z$ is given by multiplying $\mathrm{d}V_\mathrm{c}$ with the comoving number density of GCs, $n_\mathrm{GC}\left(z\right)$. We follow \cite{Fragione:2017blf, Kovetz:2018vly} and model $n_\mathrm{GC}\left(z\right)$ as\footnote{We note that assuming a constant comoving number density of GCs results in very similar results.} $n_\mathrm{GC,0}\left(H\left(z\right)/H_0\right)^3$, where $H\left(z\right)$ is the Hubble parameter and $n_\mathrm{GC,0}=3\mathrm{Mpc}^{-3}$ is the local number density of GCs~\citep{PortegiesZwart:1999nm}.
Combining Eq.~\eqref{dN per GC at detector} with these quantities (and assuming isotropy) gives us the detected event density function for the dynamic channel
\begin{flalign}\label{f_det(M1,M2,z)^dyn}
\nonumber &\frac{\partial^3 N_\mathrm{obs}^\mathrm{(dyn)}}{\partial M_1\partial M_2\partial z}=
\\&\qquad\qquad\frac{4\pi cT\chi^2\left(z\right)n_\mathrm{GC}\left(z\right)}{1+z}\frac{\tilde\Gamma\left(M_1,M_2,z\right)}{H\left(z\right)}P\left(M_1,M_2,z\right).
\end{flalign}

\subsection{Static channel}
In a very similar manner, we consider the contribution of the static channel to the event density function. We model the static channel IMF $f_0\left(M\right)$ according to Eq.~\eqref{SBH+NS IMF}, but we normalize this IMF to integrate to 1. This means that we are not concerned with the absolute amount of SBHs and NSs in the static channel --- only with their ratio. We then construct the static merger rate density as follows
\begin{equation}\label{R_stat}
 R_\mathrm{stat}\left(M_1,M_2\right)=R_0f_0\left(M_1\right)f_0\left(M_2\right)\left(\frac{M_2}{M_1}\right)^{\beta_2}.
\end{equation}
$R_\mathrm{stat}$ can be considered as the analogous static quantity of Eq.~\eqref{Gamma_i,j}. Here the static kernel parameter is $\beta_2$. Note that since $M_2$ is assumed to be smaller than $M_1$, $R_\mathrm{stat}$ is symmetric under the exchange of the coalescing masses. $R_0$ is another free parameter\footnote{\label{test}{ Note that $R_0$ in our model is slightly different than the one presented in \cite{Abbott:2020gyp}, where $f_0\left(M_2\right)$ was chosen to be a uniform distribution instead of the IMF. In addition, our normalization scheme is different as we normalize $f_0\left(M\right)$ to integrate to 1.}} in our model and it has units of $\mathrm{Gpc}^{-3}\mathrm{year}^{-1}$. It has a similar role to the role $R_{11}$ has in the dynamic channel, it controls the total amount of detected events from the static channel.

The detected event density function for the static channel is then given by
\begin{flalign}\label{f_det(M1,M2,z)^stat}
\nonumber &\frac{\partial^3 N_\mathrm{obs}^\mathrm{(stat)}}{\partial M_1\partial M_2\partial z}=
\\&\qquad\qquad\frac{4\pi cT\chi^2\left(z\right)\left(1+z\right)^3}{1+z}\frac{R_\mathrm{stat}\left(M_1,M_2\right)}{H\left(z\right)}P\left(M_1,M_2,z\right).
\end{flalign}
Note that in the numerator of Eq.~\eqref{f_det(M1,M2,z)^stat} we have inserted the factor $\left(1+z\right)^3$,  reflecting our assumption that the mergers within the static channel have a constant comoving density.

\subsection{Combining the channels}
We use a mixing parameter $X$ between the two merger channels \citep{Baibhav:2020xdf}
\begin{flalign}\label{f_det(M1,M2,z)}
\nonumber &f_\mathrm{det}\left(M_1,M_2,z\right)\equiv\frac{\partial^3 N_\mathrm{obs}}{\partial M_1\partial M_2\partial z}
\\&\qquad\quad=X\frac{\partial^3 N_\mathrm{obs}^\mathrm{(dyn)}}{\partial M_1\partial M_2\partial z}+\left(1-X\right)\frac{\partial^3 N_\mathrm{obs}^\mathrm{(stat)}}{\partial M_1\partial M_2\partial z}.
\end{flalign}
Thus, $X=1$ ($X=0$) corresponds to contributions from only the dynamic (static) channel.

Eqs.~\eqref{coagulation equation} and \eqref{f_det(M1,M2,z)} are the most important equations in this work. While Eq.~\eqref{coagulation equation} describes how the dynamic channel evolves, Eq.~\eqref{f_det(M1,M2,z)} predicts the number of detected events for each possible combination of the three independent variables $M_1$, $M_2$ and $z$. By marginalizing $f_\mathrm{det}\left(M_1,M_2,z\right)$ over two of these variables we get the event distribution with respect to the third. We can also make the transformation $\left(M_1,M_2\right)\to\left(q,M_\mathrm{sum}\right)$ where $q$ is the mass ratio and $M_\mathrm{sum}=M_1+M_2$, which will then allow us to calculate the event distribution with respect to either $q$ or $M_\mathrm{sum}$.

Marginalizing over (any) three variables gives the total number of events for a particular experiment with observation time $T$ and probability curve $P\left(M_1,M_2,z\right)$. We calculate the probability curve using the code of \cite{Chen:2017wpg}. We calculate it for both aLIGO's O3a run  ($T\sim177.3\,\mathrm{days}$ for O3a \citep{Abbott:2020niy}) and for the future aLIGO O5 run ($T\sim0.8\cdot6=4.8\,\mathrm{yrs}$, where a duty cycle of 80\% was assumed). To do so, we rely on aLIGO's O3a and O5 noise curves which are given in \cite{Abbott:2020niy, Aasi:2013wya}. As demonstrated by Fig.~\ref{figure_8}, the probability to detect any merger would be much greater for O5, thanks to its predicted improved sensitivity.

Finally, we also account for errors in the estimated masses from the strain signal. For that matter, we simplistically assume that the errors are normally distributed around zero with a standard deviation $\sigma$.
In other words, if we define $x\equiv M/M'>0$, where $M$ is the detected mass and $M'$ is the true mass, then $x$ is distributed as\footnote{The normalization of $f_G\left(x\right)$ is determined by requiring the same total number of events before and after it was applied.}
\begin{equation}\label{smoothing Gaussian}
f_{\mathrm{G}}\left(x\right)\propto\mathrm{exp}\left(-\frac{\left(x-1\right)^{2}}{2\sigma^{2}}\right)\mathcal{H}\left(x\right),
\end{equation}
where $\mathcal{H}\left(x\right)$ is the Heaviside function. Suppose now that we have marginalized $f_\mathrm{det}\left(M_1,M_2,z\right)$ and obtained $f_\mathrm{det}\left(M\right)$ (where $M$ is either $M_1$ or $M_2$). After the last correction, the density function is then given by
\begin{eqnarray}\label{smoothing}
\nonumber f_{\mathrm{obs}}\left(M\right)&&=\iint f_{\mathrm{det}}\left(M'\right)f_{\mathrm{G}}\left(x\right)\delta\left(M-xM'\right)\mathrm{d}x\mathrm{d}M'
\\&&=\int_{M'=0}^{\infty}f_{\mathrm{det}}\left(M'\right)f_{\mathrm{G}}\left(\frac{M}{M'}\right)\frac{\mathrm{d}M'}{M'}.
\end{eqnarray}
We apply the same procedure for the $q$ and $z$ distributions as well. As for the value of $\sigma$, for our O3a simulations we adopt the median relative 1-$\sigma$ uncertainty from GTWC-2. The values are $\sigma_{M_1}\approx24\%$, $\sigma_{M_2}\approx27\%$, $\sigma_{z}\approx40\%$. For our O5 simulations we assume much improved sensitivity, corresponding to an uncertainty\footnote{ The optimistic assumption of lower uncertainties in the merger parameters for future experiments is justified as enhanced detectors with better sensitivity in the low-frequency regime will be able to probe many more cycles of inspiral, affecting the SNR.} of $\sigma_{M_1}=\sigma_{M_2}=\sigma_{z}=5\%$. In any case, the uncertainty in the $q$ distribution is assumed to be $\sigma_q=\sqrt{\sigma_{M_1}^2+\sigma_{M_2}^2}$.

\section{Observable features of the dynamic channel}\label{Comparison to observations}

In this section we compare our model predictions with the O3a results. We primarily focus on the dynamic channel ($X=1$). In addition, we also plot the distribution with respect to the static channel ($X=0$) for a $\beta_2=2$ kernel via a black curve. In the next section we allow some mixing between the channels and let $X$ vary. In all cases we set the value of $R_{11}$ and $R_0$ to match the total number of events (39).\footnote{ For the static channel, we have verified that we gain $\sim39$ events by adopting the best-fit values from \cite{Abbott:2020gyp} with the adequate normalization method (see footnote~\ref{test}).}

For comparison purposes, we present O3a data in gray histograms which follow the PDFs of the variables. These PDFs were constructed by combining all the PDFs of the 39 events together. The PDFs of $M_1$, $M_2$ and $z$ for each event were approximated by asymmetrical Gaussians centered at the best fit values. We then randomly sampled the $M_1$ and $M_2$ PDFs of each event thousands of times, and from these samples constructed the $q$ PDF.

\subsection{Mass ratio}

\begin{figure}
\includegraphics[width=\columnwidth]{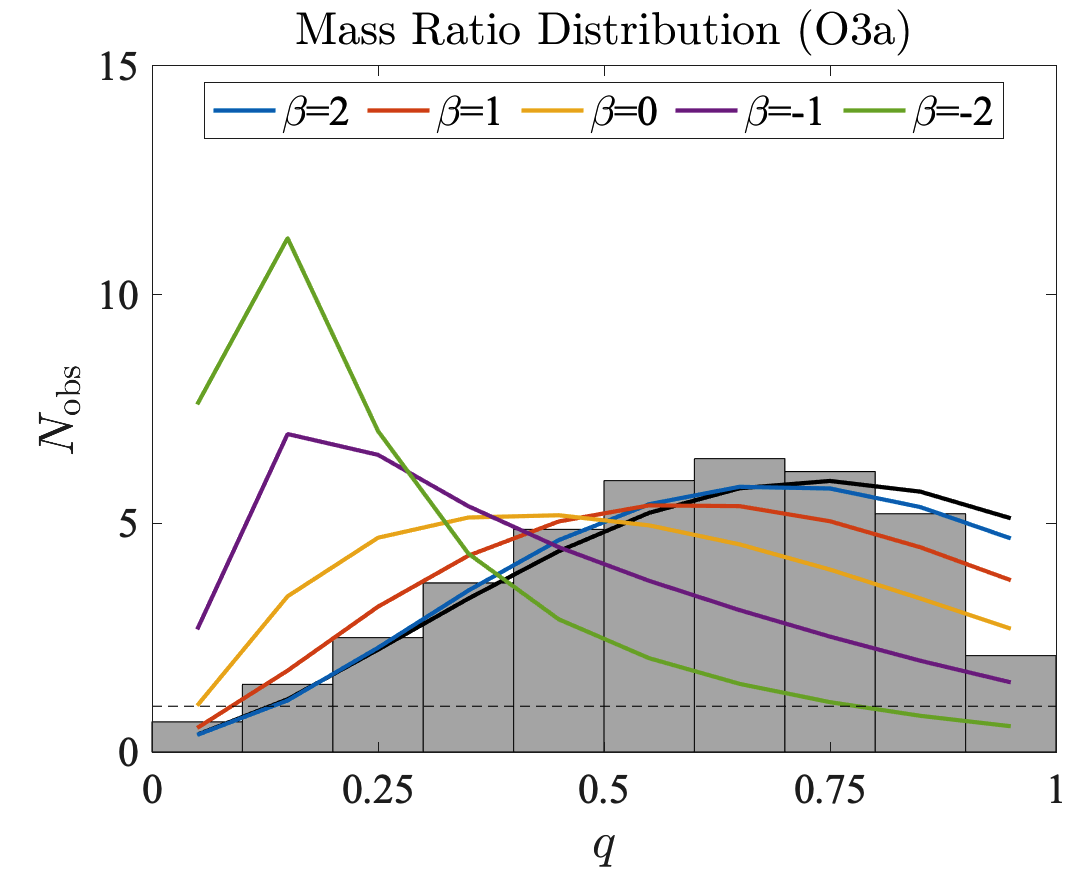}
\caption{Expected distribution for the mass-ratio parameter $q=M_2/M_1$ for different $\beta$ values when only the dynamic channel is considered ($X=1$).
In all cases we set $\gamma=0$, $N_\mathrm{0,SBH}=300$, $N_\mathrm{0,NS}=0$, no ejections nor mass loss and zero delay time.
The black curve corresponds to the static channel ($X=0$) with $\beta_2\!=\!2$. The histograms correspond to O3a data. For each curve, the overall rate parameters $R_{11}$ and $R_{0}$ were chosen to match the total number of events in O3a (39).}
\label{figure_9}
\end{figure}

We first assume no NS contribution, i.e.~we take Eq.~\eqref{stellar IMF} as the IMF, and we assume $N_\mathrm{0,SBH}=300$ with no ejections nor mass loss and zero delay time. Since the $\beta$ index is associated with only the quotient term of the kernel, it mostly affects the event distributions with respect to $q$, while $\gamma$ mostly affects the $M_\mathrm{sum}$ distribution. In Fig.~\ref{figure_9} we plot the event distribution with respect to $q$ for different $\beta$ values. 

The constant-kernel case of $\beta=0$ has no preference for any kind of merger, yet it does not yield a flat distribution for $q$ and it peaks at $q\sim0.5$.
This is due to a number of reasons: 
First, the cluster BHMF is not flat and there are  more light BHs than heavy BHs (cf.\ Fig.~\ref{figure_2}). Secondly, the lightest and heaviest BHs in the GC have finite masses, which prevents mergers with $q\to0$. Thirdly, at fixed total mass aLIGO is more sensitive to mergers with higher $q$ as the emitted GW energy is greater for such mergers (cf.\ Eq.~\eqref{E_GW}). Therefore, the $\beta=0$ curve in Fig.~\ref{figure_9} decreases towards the small $q$ regime. On the other hand, $q\approx1$ events are less favored due to the large uncertainty in measurements --- this uncertainty (reflected in our analysis by Eq.~\eqref{smoothing}) results in smoothing the distribution, thus migrating actual $q\approx1$ events to lower values of $q$. The peak of the distribution shifts further to higher $q$ for positive $\beta$ values which generate more mergers with equal masses. Negative $\beta$ values, conversely, clearly exhibit the opposite trend. The distributions in these cases suggest that most of the detected events have small $q$ with almost no events toward $q\approx1$.

Clearly, the O3a data fits better to the positive $\beta$ values, while negative $\beta$ values can be ruled out as the sole contributor to BH mergers, in accordance with \cite{Fishbach:2019bbm}. Despite the fact that kernels with positive $\beta$ strongly prefer equal masses, they still allow events such as GW190412~\citep{LIGOScientific:2020stg} and GW190814~\citep{Abbott:2020khf} which have estimated mass ratios of $q\sim0.28$ and $0.11$, respectively. Notice the similarity between the black curve which corresponds to the static channel with $\beta_2=2$ and the blue curve which corresponds to the dynamic channel with $\beta=2$. If one demands a certain total number of events, the shape of the distribution does not change significantly, even when other values of $\gamma$ and ejections are considered. We therefore adopt $\beta=\beta_2=2$ in the remaining parts of this work. This choice is consistent with \cite{Abbott:2020gyp} where only the static channel was considered.

\subsection{Upper mass gap}
\begin{figure}
\includegraphics[width=\columnwidth]{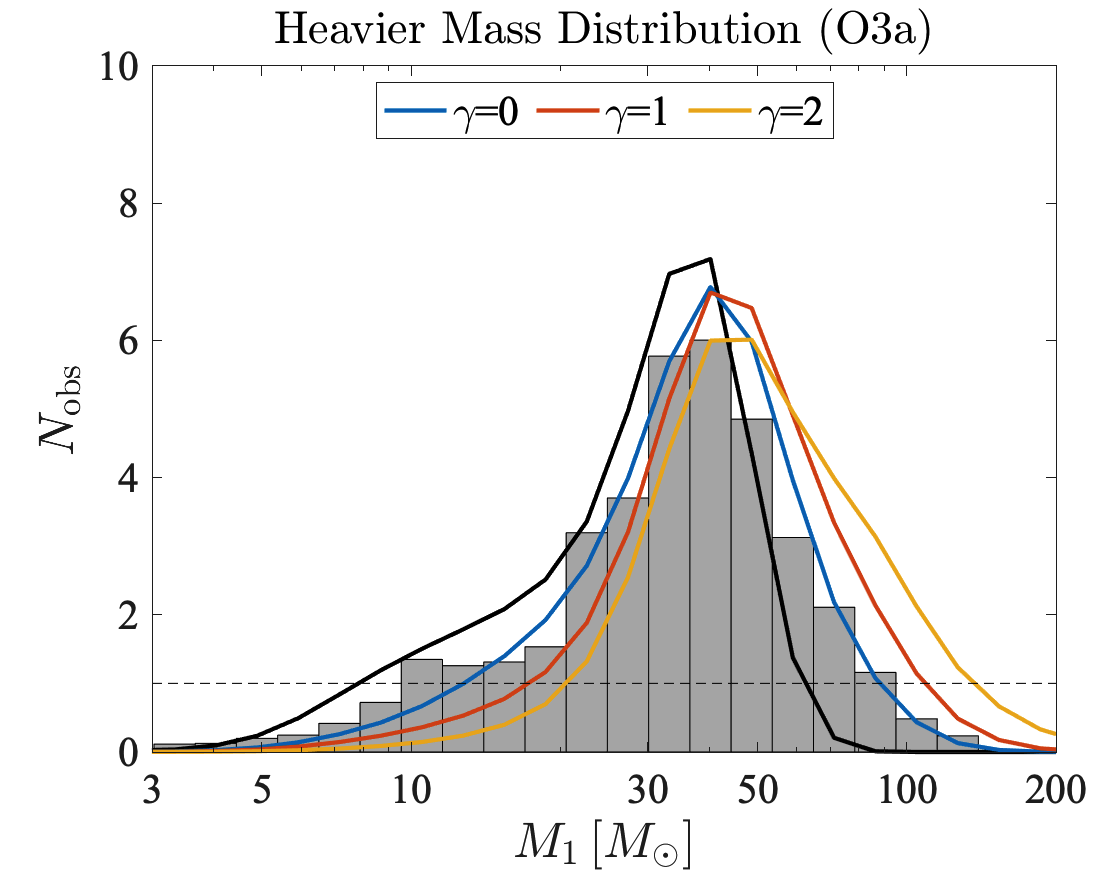}
\caption{Distributions for $M_1$ ($>M_2$) for different $\gamma$ values, in all cases with $\beta=2$, but otherwise with the same specifications as Fig.~\ref{figure_9}.
}
\label{figure_10}
\end{figure}

If only the dynamic channel is considered then the value of $\gamma$ determines the largest $M_1$ and $M_2$ to be detected. Fig.~\ref{figure_10} presents the expected $M_1$ distribution for different values of $\gamma$. As $\gamma$ becomes larger, there are more events with heavier BHs, at the expense of events with lighter BHs. 
Due to the coagulation process, our model predicts several events above the upper mass cutoff ($50M_\odot$) even when $\gamma=0$. 
As expected, it is very easy to breach the upper mass gap via BH coagulation.

\begin{figure}
\includegraphics[width=\columnwidth]{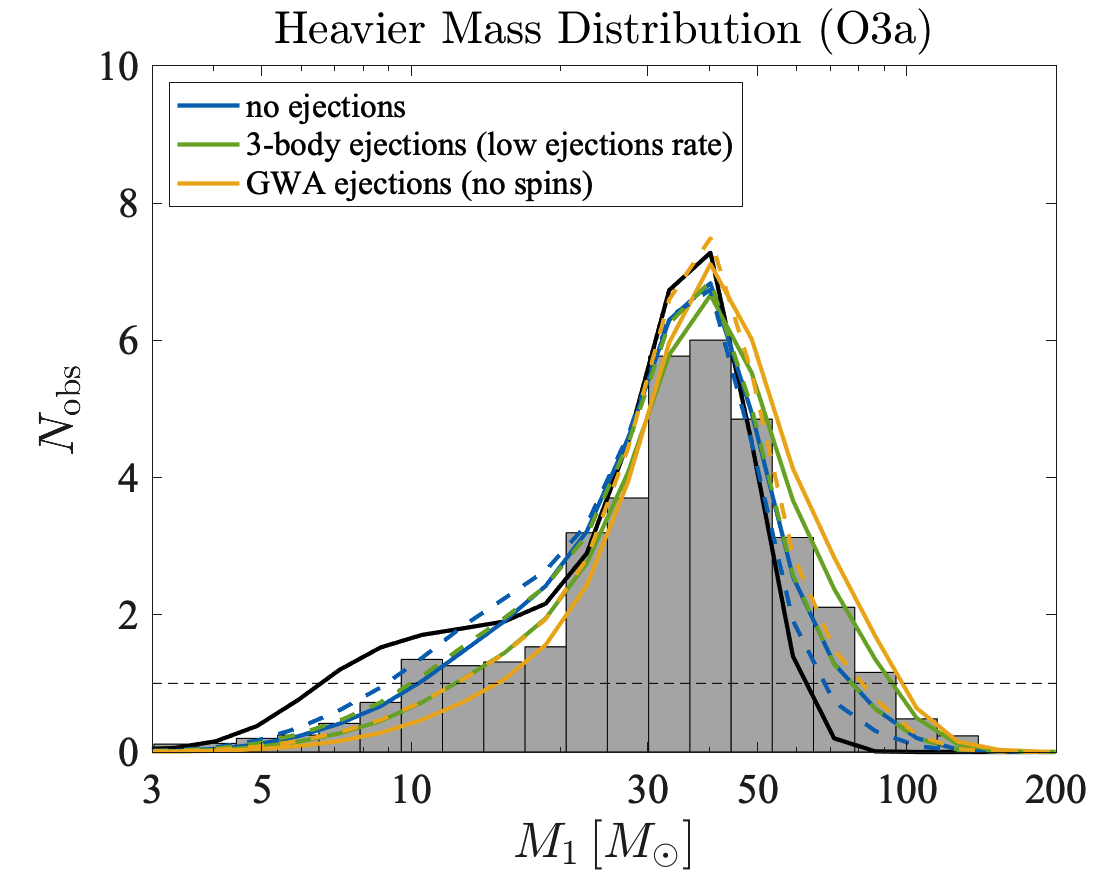}
\includegraphics[width=\columnwidth]{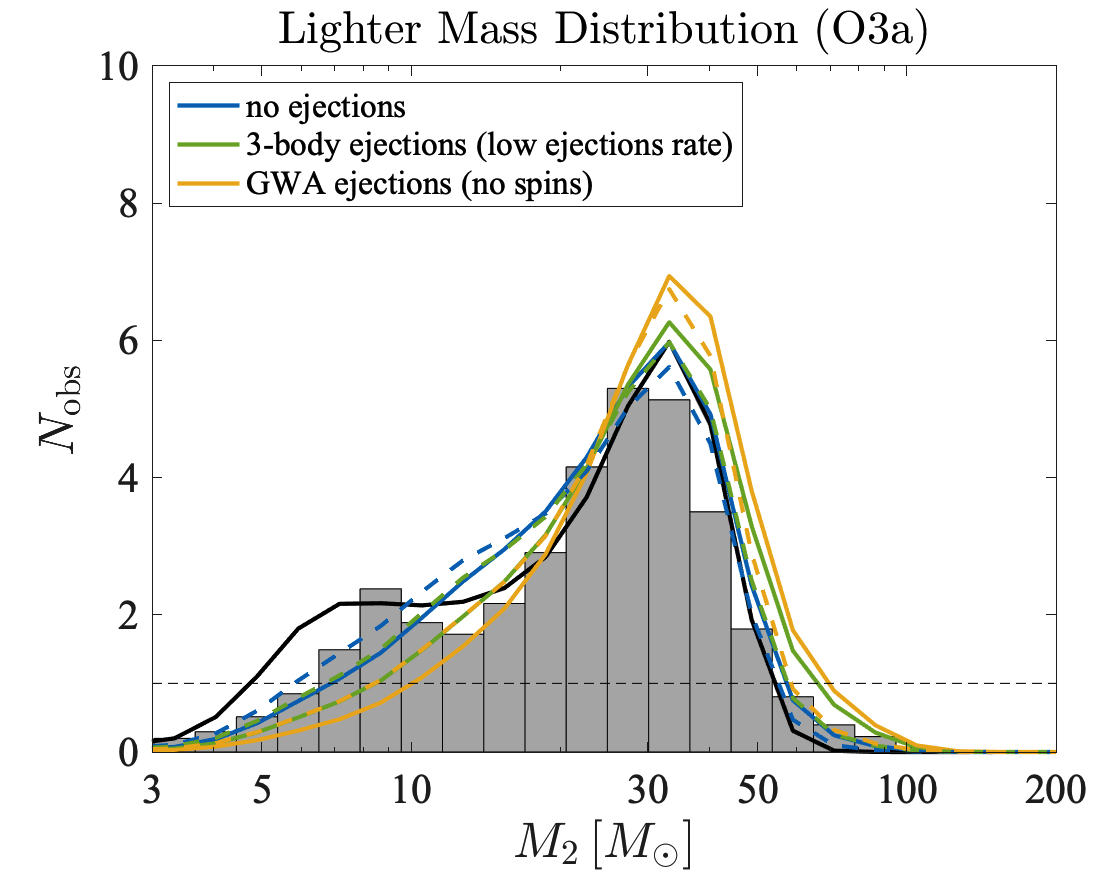}
\includegraphics[width=\columnwidth]{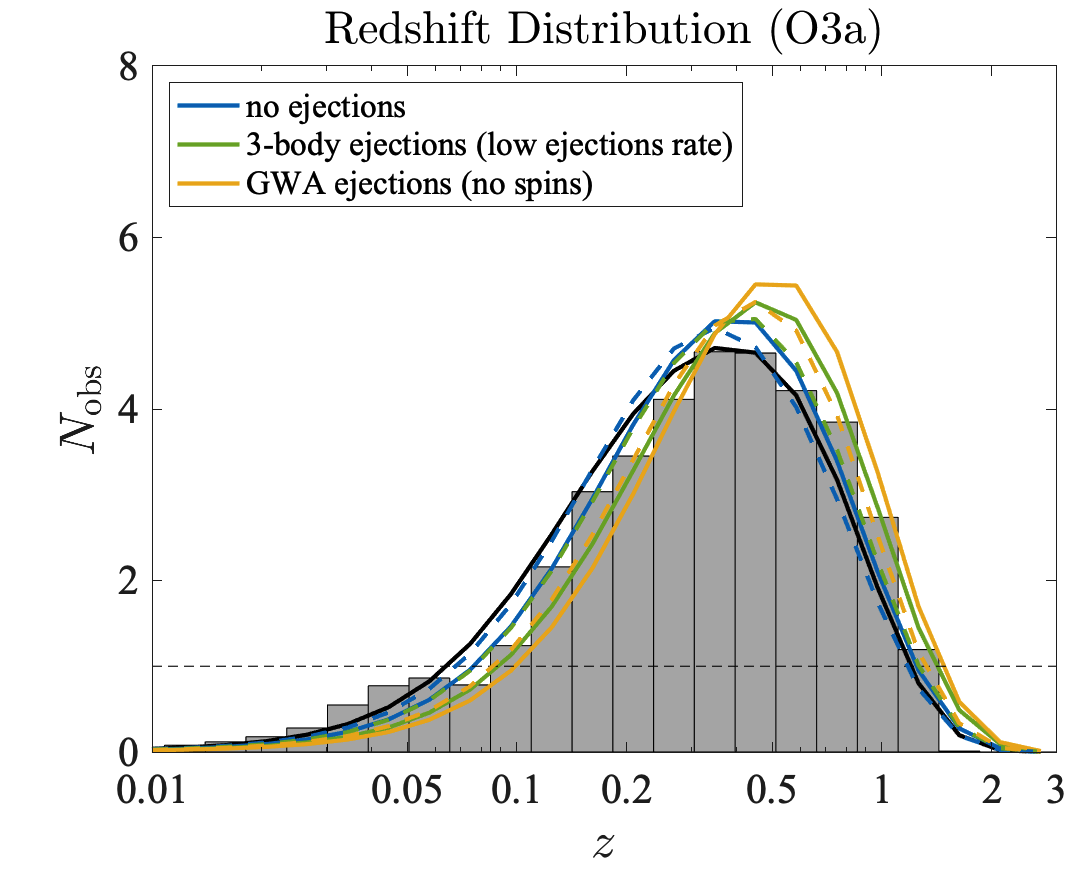}
\caption{The detected merger distribution with respect to $M_1$, $M_2$ and $z$ predicted for O3a, as in Fig.~\ref{figure_9} assuming $\beta=2$, $\gamma=0$. The same ejections parameters of Fig.~\ref{figure_3} were used here. The black curves correspond to $\beta_2=2$ static channel. Solid (dashed) curves correspond to coagulation without (with) delay time effect. For each curve, the overall rate parameters $R_{11}$ and $R_0$ were chosen to match the total number of events in O3a (39).
}
\label{figure_11}
\end{figure}

\subsection{BHMF shape}
It is also interesting to examine how our coagulation model affects the $M_1$, $M_2$ and $z$ distributions. We plot the expected distributions for $\beta=2$, $\gamma=0$ kernel in Fig.~\ref{figure_11}. As before, the black curves in these figures correspond to the expected distributions when coagulation effects are not included (i.e.\ only the static channel is considered with $X=0$). Without coagulation, most of the detected events have $M_1,M_2\sim40M_\odot$ (due to the PPSN peak and the biased sensitivity) or $M_2\sim5M_\odot$ (due to the abundance of light BHs, cf.\ Fig.~\ref{figure_2}), and there are almost no events with $M_1$ either below $5M_\odot$ or above $50M_\odot$ (the small fraction of events in these regions are due to the large uncertainty in measurements).

We show in Fig.~\ref{figure_11} the distributions when solving the coagulation equation. 
We see how, when ejections are not included, as light BHs merge, their abundance decreases, and they become less detectable today. 
As a consequence of mergers, however, heavier BHs are formed and breach the upper mass cutoff at $M\approx 50 M_\odot$.
We also show in Fig.~\ref{figure_11} the result when modelling ejections, either 3-body ejections with $M_\mathrm{esc}=5M_\odot$ or GWA ejections with \linebreak $v_\mathrm{esc}=50~\mathrm{km/s}$ (see Fig.~\ref{figure_3}).
Since there are fewer light BHs due to their ejection, the interaction rate ($R_{11}$) ought to be increased to reach the O3a number of observed events (39).
As a result, the retained BHs merge more frequently, and so the observed distribution contains more heavy BHs, whose mergers are more readily detectable.

The dashed curves of Fig.~\ref{figure_11} correspond to the same ejections scenarios discussed previously, but now the delay time distribution is introduced. When the cluster is free of ejections, the delay time causes less merger events. This may seem at first to contradict the results of Fig.~\ref{figure_7}, where the final total merger rate is greater when the delay time is introduced. Yet, one has to keep in mind that the total amount of observed merger events directly depends on the detection probability $P\left(M_1,M_2,z\right)$ which is very small for small $M_1$, $M_2$ values. Because the delay time delays the formation of heavier BHs, the mergers in the delay time scenario are less detectable, even though the merger rate is greater. In order to obtain the desired 39 events, $R_{11}$ must be (slightly) increased and this consequently reduces the impact of the delay time effect. Eventually, the net effect of the delay time in our model is to increase the detected small mass merger events at the expense of heavy mass merger events. This effect is more pronounced when ejections are active, because ejections are not delayed in our model. Nevertheless, we learn that the delay time effect does not alter much the observed BHMF (even if we consider a mass dependent $t_{d,\mathrm{min}}$ --- see details at appendix \ref{Mass dependent td,min}), and for that reason we do not include it in the rest of this work (see discussion in section \ref{Conclusions}).

As can be seen by the lower panel of Fig.~\ref{figure_11}, our model predicts that the distribution of mergers peaks around $z\sim0.5$.
There is a balance between the larger number of GCs that can be reached up to higher $z$, and the lower sensitivity of aLIGO to events from farther distances.
These two factors are reflected in our model by  $\chi\left(z\right)$, $n_\mathrm{GC}\left(z\right)$, and  $P\left(M_1,M_2,z\right)$ (cf. Eq.~\eqref{f_det(M1,M2,z)^dyn}). When heavier BHs are detected, the $z$ distribution is shifted to higher $z$, indicating that we have more chance to detect these BHs at farther distances, as they are more luminous.

\begin{figure*}
\includegraphics[width=\columnwidth]{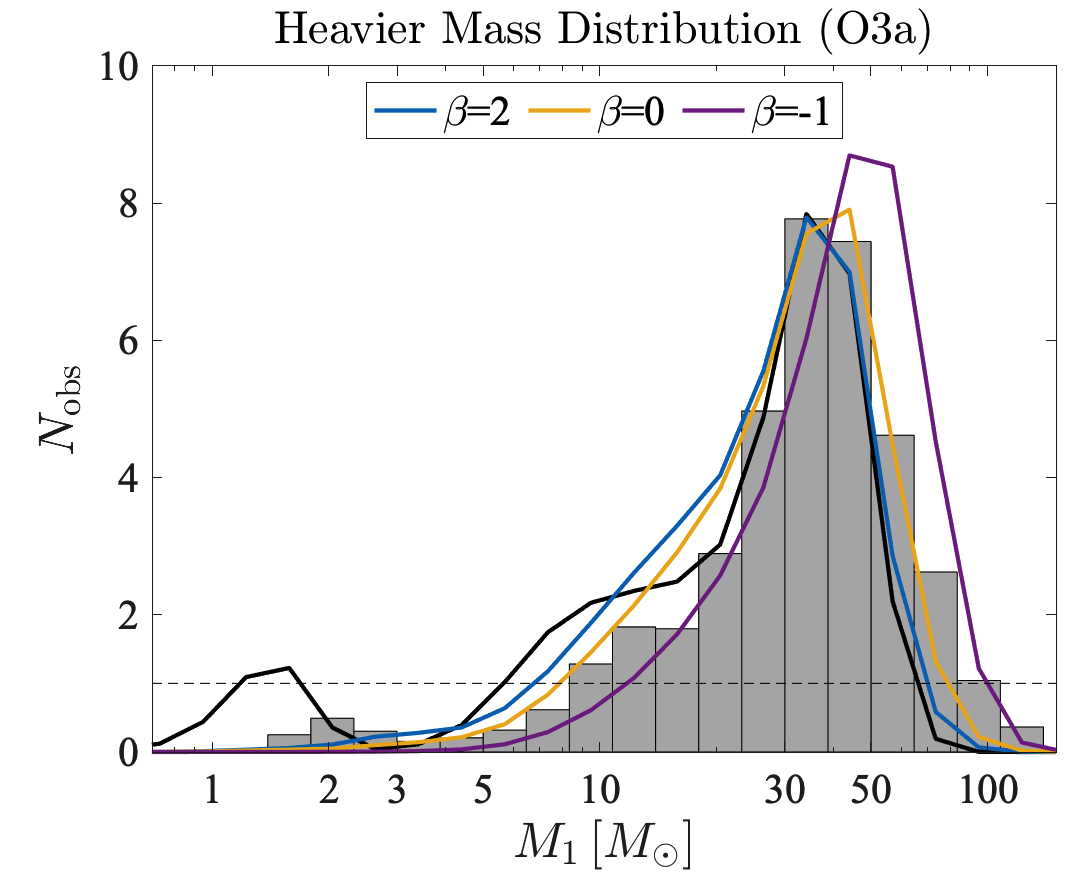}
\includegraphics[width=\columnwidth]{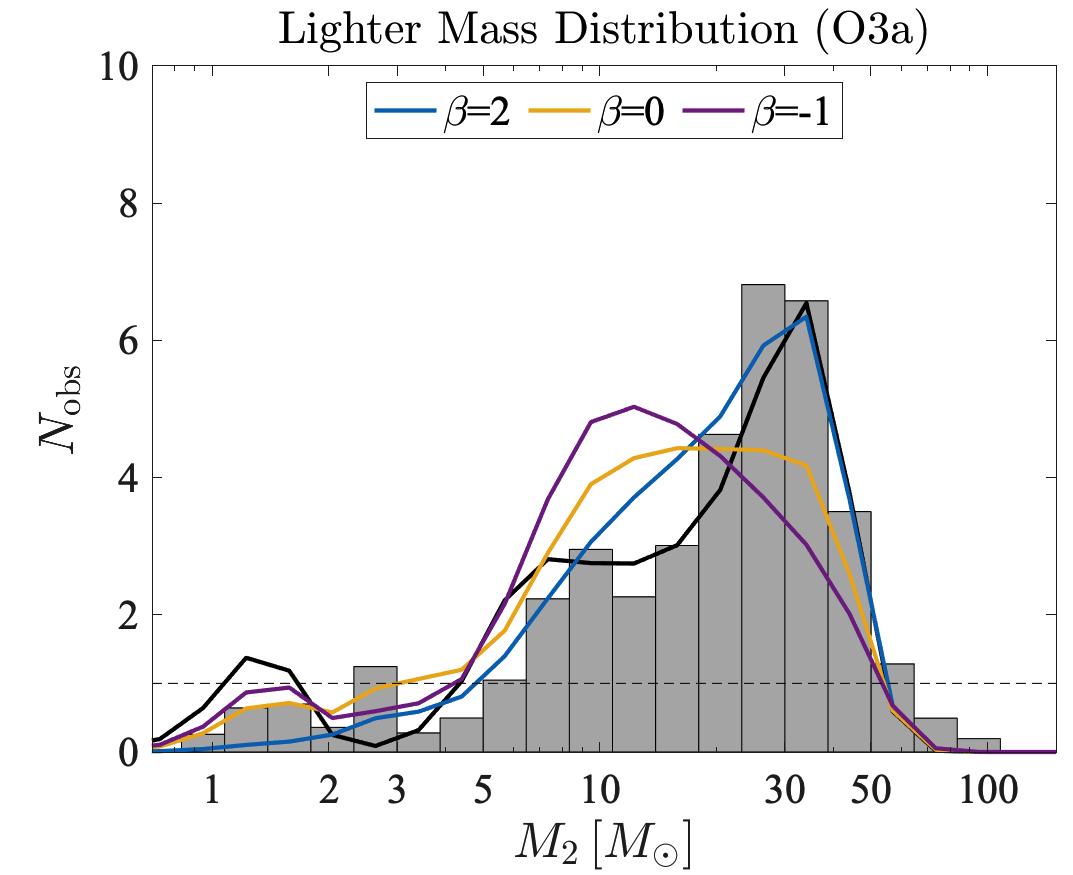}
\caption{The detected merger events distribution for different values of $\beta$ with respect to $M_1$ and $M_2$ as predicted for the O3a aLIGO run, as in Fig.~\ref{figure_9}, but with with $N_\mathrm{0,NS}=1000$ NSs. 
As before, black curves correspond to the static channel ($\beta_2=2$).}
\label{figure_12}
\end{figure*}

\subsection{Lower mass gap}
Since SBHs are considered to have a minimum mass of $M_\mathrm{min}\!=\!5M_\odot$ (see Eq.~\eqref{Ely's IMF}), it is impossible to achieve events in the lower mass gap with only coagulation of SBHs. The GW190814 event had a smaller component mass of $M_2\!\sim\!2.6M_\odot$, 
 indicating it may have originated from the coalescence of two NSs of mass $1.3M_\odot$ (cf.\ Fig.~\ref{figure_5}). 
It is therefore tempting to test whether our model can generate an event akin to GW190814 (as well as any other events with detected components in the lower mass gap) via NS coagulation. 
We do so now.

Naively, we can set $N_\mathrm{0,NS}/N_\mathrm{0,SBH}\approx4$, which is consistent with a Salpeter power-law of $\alpha=2.35$ for the BH and NS stellar progenitors. Ejections of NS from the cluster due to their natal kicks can strongly affect this ratio~\citep{Ye:2019xvf}. Yet, we find that if the initial NSs-SBHs ratio is of order unity (or less), then it is unlikely to yield events at the lower mass gap if one demands 39 events in total. We therefore consider clusters with $N_\mathrm{0,NS}>N_\mathrm{0,SBH}$.

By assuming the IMF to be of the form of Eq.~\eqref{SBH+NS IMF}, with $N_\mathrm{0,SBH}=300$ and $N_\mathrm{0,NS}=1000$ NSs, solving the coagulation equation with $\beta=\gamma=0$, and then applying Eq.~\eqref{f_det(M1,M2,z)} to calculate the distribution of detected events, we find it easy to explain the smaller mass component of GW190814 as a product of NS coagulation in a GC\footnote{We note, however, that simulations of the dynamics of BHs and NSs within GCs cores predict a smaller mass segregation for NSs, which leads to negligible merger rates~\citep{Ye:2019xvf}.}.

\begin{figure}
\includegraphics[width=\columnwidth]{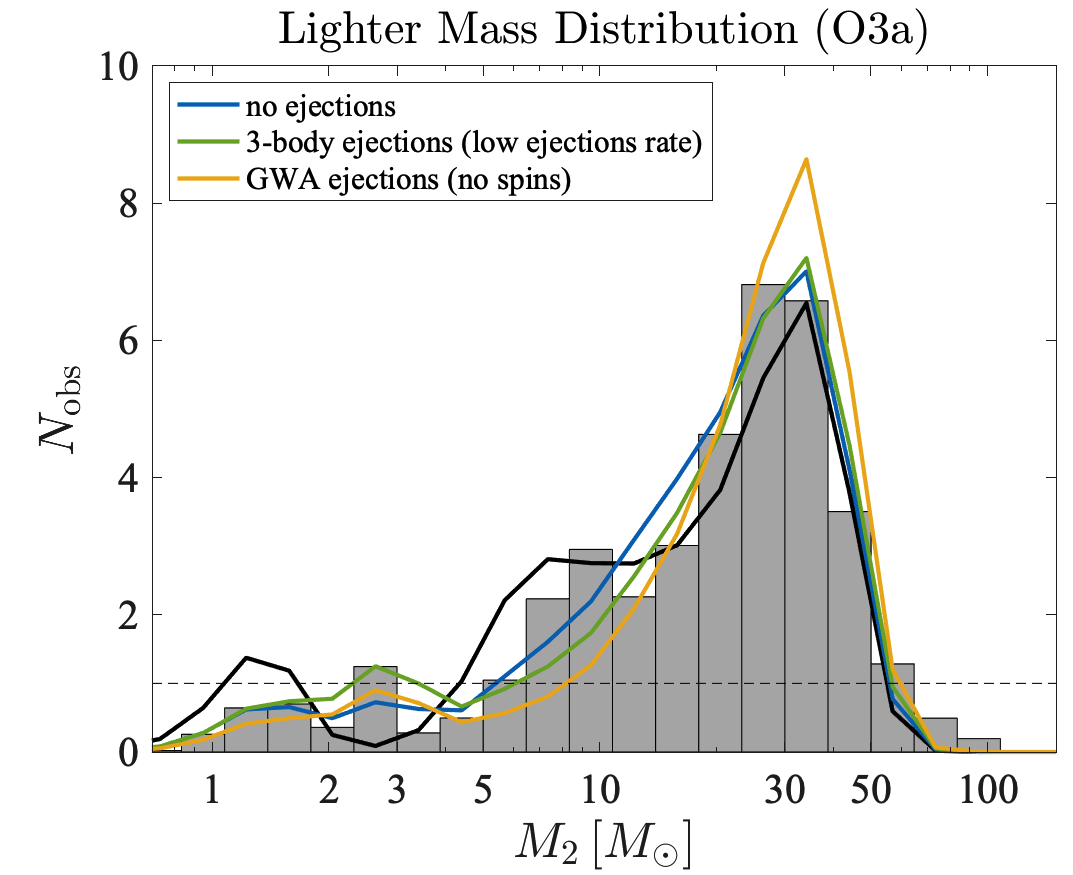}
\caption{
Same as Fig.~\ref{figure_12}, albeit the NSs are kept frozen for $10\mathrm{Gyrs}$, to account for a possible delay in their sinking to the GC core. The same ejections parameters of Fig.~\ref{figure_3} were used here.
The black curve corresponds to $\beta_2=2$ static channel as before.}
\label{figure_13}
\end{figure}

We show the result of this calculation in
Fig.~\ref{figure_12}. 
As before, the black curves correspond to the static channel (with $\beta_2=2$). 
We see that, on its own, the static channel contributes $\sim2$ merger events that involve NSs. From the distribution of $M_1$ we learn that for this kernel these NS mergers are likely to involve 2 NSs (rather than one NS and one BH).
When we solve the coagulation equation we find it harder to detect NSs, as their continuous mergers deplete their abundance by today.
Nevertheless, each NS-NS merger results in a $\sim2.6M_\odot$ BH, whose subsequent merger can be detected.
Our model predicts that O3a should have found $\sim\!1$ of these mass-gap events, roughly independently of $\beta$, although the rate is higher for $\beta=0$. The reason is that positive $\beta$ prefers equal-mass mergers, whereas negative $\beta$ leads to runaway depletion in the lighter objects (cf.\ Fig.~\ref{figure_2}).

\begin{figure*}
\includegraphics[width=\columnwidth]{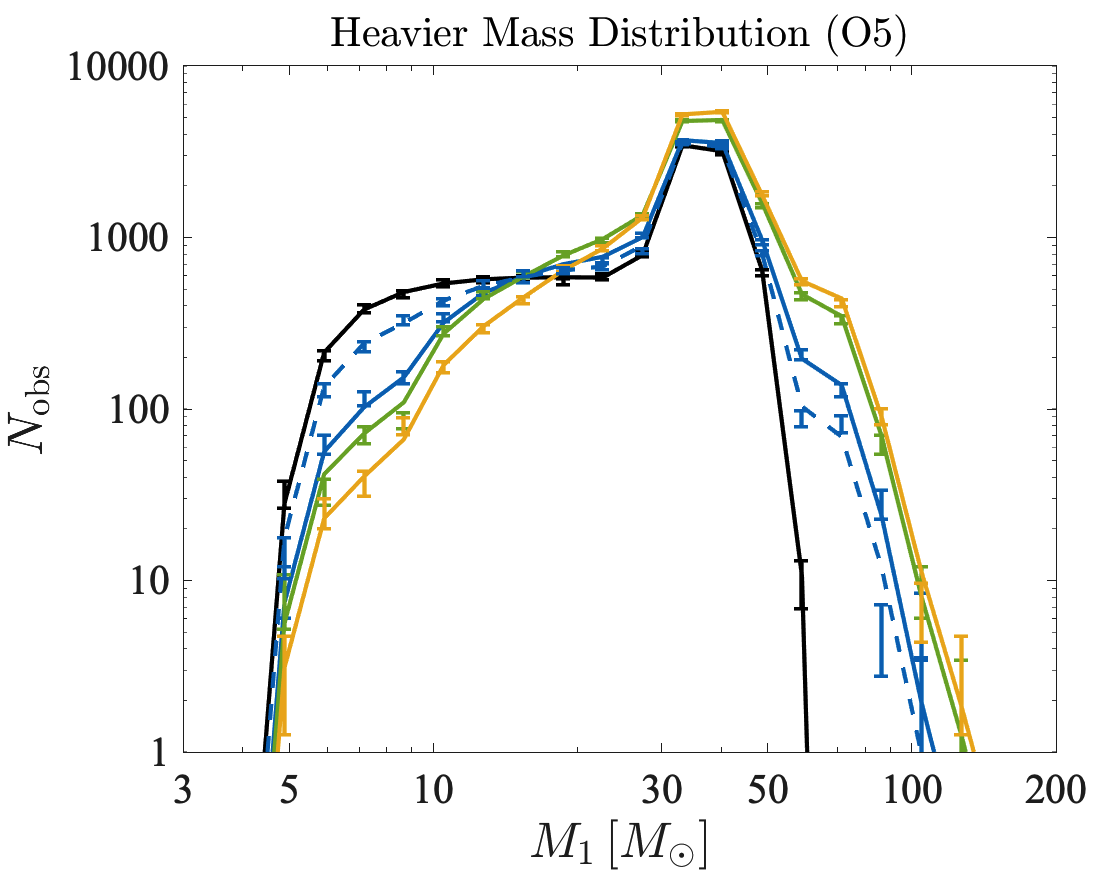}
\includegraphics[width=\columnwidth]{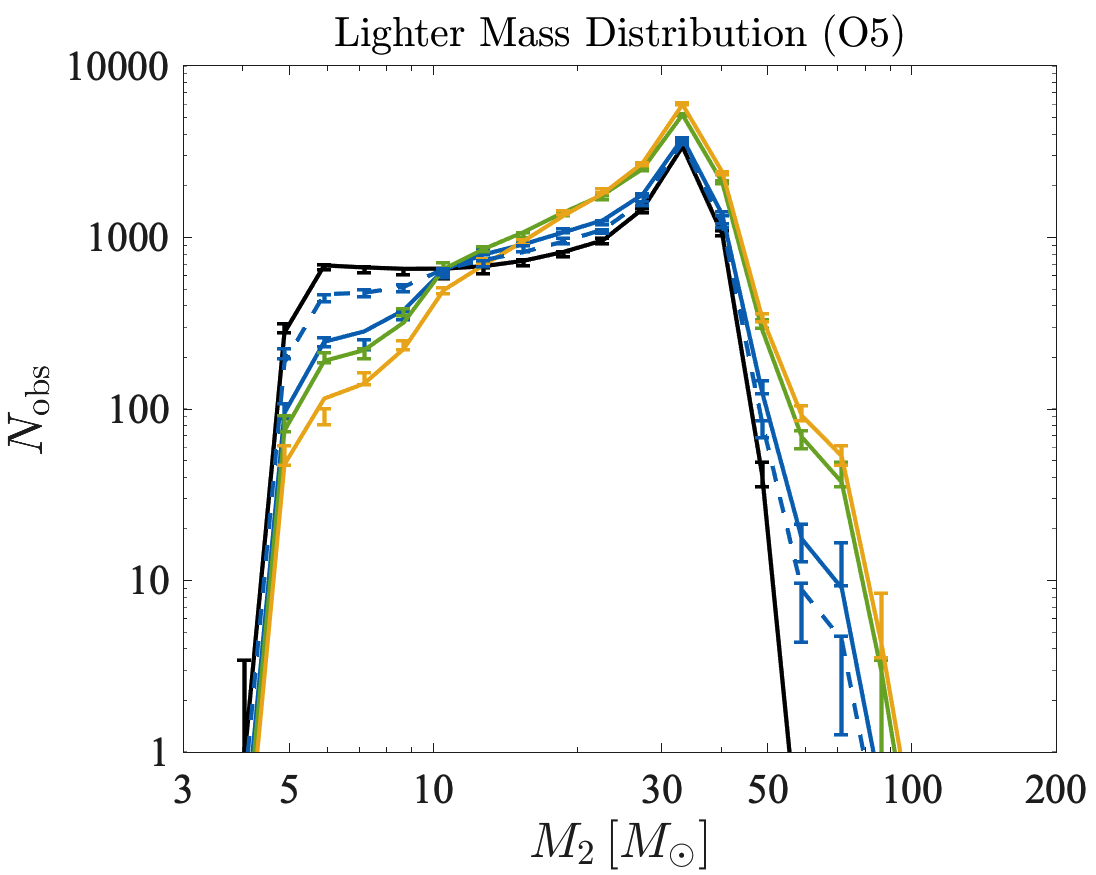}
\caption{The predicted distribution of mergers for  O5, with $\beta=\beta_2=2$, $\gamma=0$, $N_\mathrm{0,SBH}=300$, $N_\mathrm{0,NS}=0$, $f_\mathrm{loss}=0$. Blue, red and green curves assume only dynamical binaries ($X=1$), with the same ejections parameters as in Figs.~\ref{figure_11} and \ref{figure_13}. The blue dashed curves correspond to an equal mix of dynamical and static binaries ($X=0.5$), with $M_\mathrm{esc}=v_\mathrm{esc}=\infty$ (no ejections). The black curves correspond to the static channel (for $X=0$). The center of each error bar was randomly selected via a Poisson distribution with a parameter that equals to the prediction of our model. For each curve the value of $R_{11}$ was chosen to match the number of events detected in O3a (yielding, for instance, $\sim12,\!400$ events for the blue curve).}
\label{figure_14}
\end{figure*}

Since NSs are much lighter than BHs, it may take more time for them to sink towards the GC center due to dynamical friction, which can produce a ``delay" in their mergers. 
We study the impact that such a delay would have on our results in Fig.~\ref{figure_13}, where we introduce the NS IMF to the simulation only after $10^{10}$ yr. 
In this case it is easier to detect NS mergers, as now depletion of NSs, via mergers, starts to take place much later in the GC lifetime. 
Moreover, we also test how our results change when modeling ejections, with the ejections rates we used in Fig.~\ref{figure_11}.
As before, including ejections forces a higher merger rate $R_{11}$ in order to achieve 39 events in O3a. 
As a result, the $M_2$ distribution is changed in Fig.~\ref{figure_13}. 
This reduces the detectability of NS mergers, as most NS mergers happen at early times (cf. Fig.~\ref{figure_6}) beyond the reach of aLIGO. 
As was noted in Fig~\ref{figure_11}, the rapid merger rate generates more SBHs above $\sim40M_\odot$, and those are more readily detected.

Finally, we note that there is another possibility to gain events at the lower mass gap via our model. In all previous cases we set $M_\mathrm{min}=5M_\odot$ as the cutoff below where no SBHs are formed. 
However, this limit is primarily motivated from empirical aspects rather than theoretical ones, namely the lack of detections of BHs below $5M_\odot$ via their X-ray emission. 
This could be the result of a real mass gap between BHs and NSs, or perhaps a poorly understood technical limitation to detect smaller BHs through this method. 
It is therefore possible that the lower mass gap does not exist at all \citep{Kreidberg:2012}, a possibility which we can study by setting $M_\mathrm{min}=2M_\odot$ (roughly the upper mass limit for NSs from both theoretical and empirical considerations~\citep{Oppenheimer:1939ne,Ozel:2016oaf}), and ignore completely the NSs contribution to the BHMF evolution. Doing so, we can indeed achieve some events at the lower mass gap\footnote{We thank Ofek Birnholtz for suggesting this investigation.}.

\section{Forecasts}\label{Forecasts}

While the O3 run has ended with dozens events in total, future experiments are expected to detect thousands of events, due to both improved sensitivity and longer observation time~\citep{Aasi:2013wya,Hild:2010id,Broeck:2013rka,2010CQGra..27s4002P,Maggiore:2019uih,Evans:2016mbw,Reitze:2019iox,Audley:2017drz}. 
In this section we focus on the planned O5 run of aLIGO. The O5 design sensitivity is by far enhanced compared to O3, which is reflected in our model by the detection probability $P\left(M_1,M_2,z\right)$ (see Fig.~\ref{figure_8}).
Moreover, the longer duration of O5 (6 years instead of $\sim1$ for O3), will allow us to probe the BHMF more deeply, as demonstrated in Fig.~\ref{figure_14} (here we use blue dashed curves to plot the result of equal mixture of the dynamic channel and the static channel, i.e.\ $X=0.5$).

\begin{figure*}
\includegraphics[width=\columnwidth]{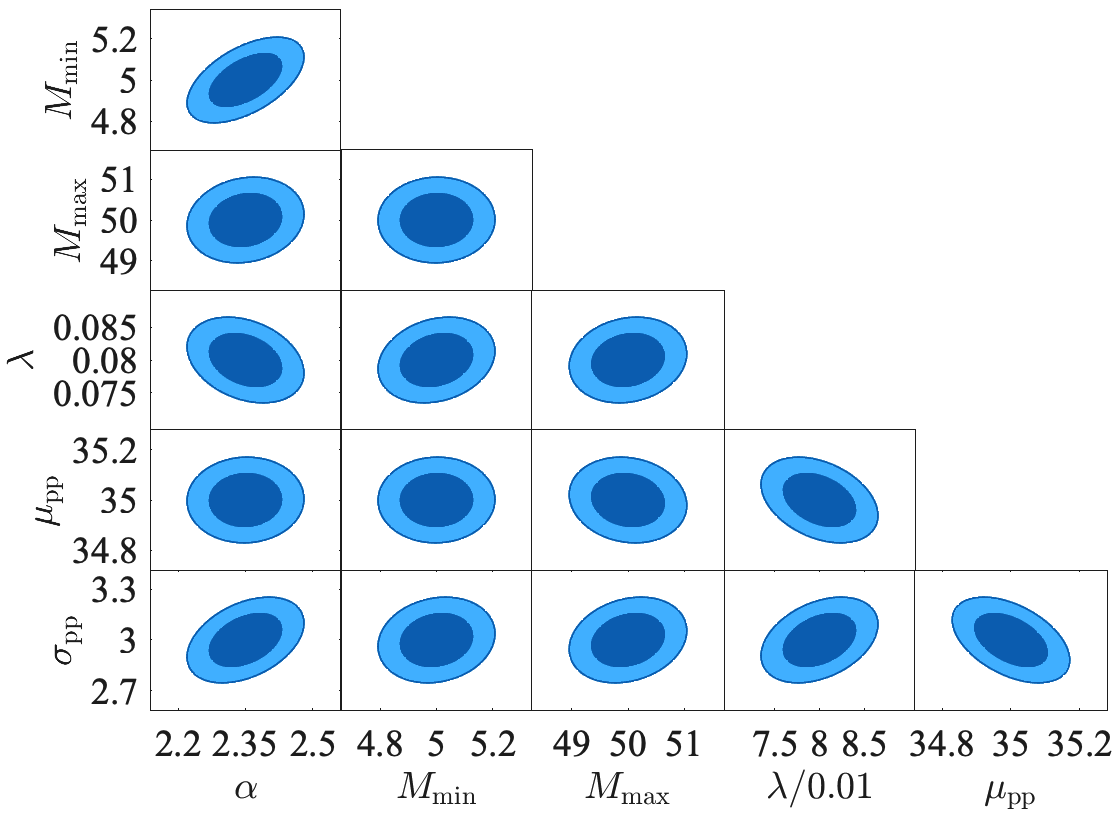}
\includegraphics[width=\columnwidth]{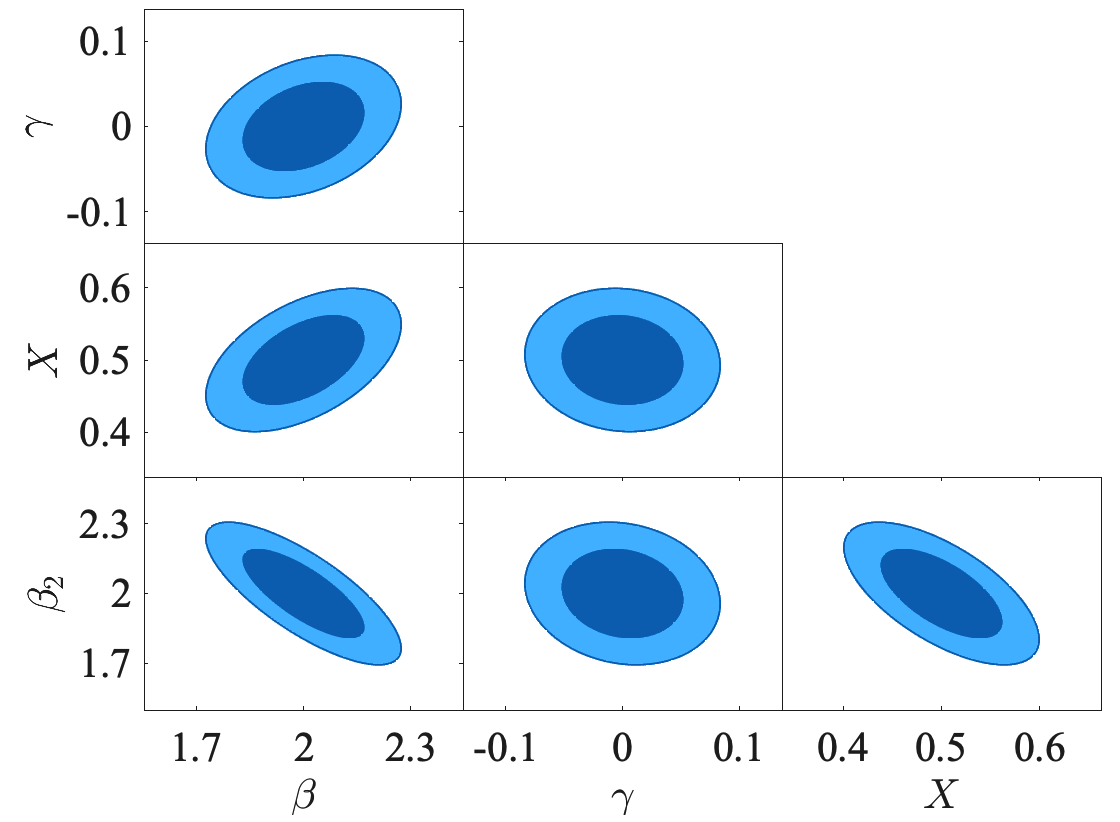}
\caption{Forecasted confidence ellipses for the parameters of our model from O5 data (with $\sim14,\!500$ merger events). Here we assumed the fiducial values $\beta=\beta_2=2$, $\gamma=0$, $N_\mathrm{0,SBH}=300$, $v_\mathrm{esc}=\,50\mathrm{km/sec}$, and $X=0.5$ (plus no NSs, fixing $N_\mathrm{0,NS}=0$). As before, the value of $R_{11}$ and $R_0$ were fixed to have 39 events for O3a, and we assumed no mass loss in mergers ($f_\mathrm{loss}=0$). }
\label{figure_15}
\end{figure*}

Not only that, but the rich O5 dataset is expected to place tight constraints on our model parameters. 
In order to forecast these, we divide the 2D distribution $f_\mathrm{det}\left(M_1,M_2\right)$ into 94 mass bins, denoted as $N_\mathrm{obs}\left(i\right)$. We assume that $N_\mathrm{obs}\left(i\right)$ are independently Poisson distributed (discarding bins with fewer than 10 events) and calculate the Fisher matrix using~\citep{Jimenez:2013mga, Jungman:1995bz}
\begin{equation}\label{Fisher matrix}
F_{\alpha,\beta}=\sum_i \frac{1}{N_\mathrm{obs}\left(i\right)}\left.\frac{\partial N_\mathrm{obs}\left(i\right)}{\partial\theta_\alpha}\frac{\partial N_\mathrm{obs}\left(i\right)}{\partial\theta_\beta}\right|_\mathrm{ML}.
\end{equation} 
Here $\theta_\alpha$ denotes the model parameters, and the derivatives are computed at the maximum likelihood (in our case at the fiducial values). Once we have the Fisher matrix at our disposal, we can invert it to obtain the covariance matrix, which we use to draw confidence ellipses in Fig.~\ref{figure_15}. In order to create these figures we varied the parameters listed in table \ref{table_1}, except for $N_\mathrm{0,NS}=0$, $t_\mathrm{c}=13\mathrm{Gyears}$ and $f_\mathrm{loss}=0$ which remained fixed. As before, the values of $R_{11}$ and $R_0$ were fixed to the values which yield 39 events for O3a (with respect to the fiducial values of the other parameters that were varied).  For this analysis, we accounted only for GWA ejections in the configuration where all the binaries have zero effective spins and used a fiducial value of $v_\mathrm{esc}=50\,\mathrm{km/sec}$.

In the left panel of Fig.~\ref{figure_15} we plot the confidence ellipses for the IMF parameters after marginalizing over the kernel parameters (right panel of Fig.~\ref{figure_15}), $N_\mathrm{0,SBH}$ and $v_\mathrm{esc}$. All the constraints are quite tight, as all can be measured at the 10\% level\footnote{We have verified that the constraints we obtained are not sensitive to our assumption of low uncertainties in O5. The results hardly change if we assume $\sigma\!=\!10\%$ (see Eq.~\eqref{smoothing Gaussian}) instead of 5\%.}. 
In addition, if a lower mass gap does exist, we learn that we would be able to pinpoint its upper bound ($M_\mathrm{min}$) with a 2-$\sigma$ CL of $\sim0.2M_\odot$. Similarly, we would be able to determine the upper mass cutoff ($M_\mathrm{max}$) with a 2-$\sigma$ CL of $\sim1M_\odot$.

In the right panel of Fig.~\ref{figure_15} we plot the confidence ellipses for the kernel parameters after marginalizing over the IMF parameters (Fig.~\ref{figure_15}, left), $N_\mathrm{0,SBH}$ and $v_\mathrm{esc}$. Here we see that O5 data will yield excellent constraints on the power-law indices $\gamma$, $\beta$ and $\beta_2$, with 2-$\sigma$ CL less than 0.5. In addition, the 2-$\sigma$ CL of $X$ is of the order of 0.1, thus allowing us to pin-point whether the detected mergers originate from different environments (such as GCs and the field), providing a powerful test of our model.

\section{Conclusions}\label{Conclusions}

In this work we studied the impact of various effects on the BH merger events which are detected by aLIGO. We considered contributions from two channels: a static channel where the merger rate is constant with time, and a dynamic channel. The analysis of the contribution from the static channel is very similar to what was done in previous studies such as \cite{Kovetz:2016kpi, Kovetz:2018vly, LIGOScientific:2018jsj}. The dynamic channel was assumed to consist of mergers that originate in the dense compact-object environments of GC cores. In order to study how the BHMF and  merger rate of the dynamic channel evolve with time we solved numerically our modified coagulation equation, Eq.~\eqref{coagulation equation}, for different scenarios. Then, by accounting for the aLIGO observational bias we were able to predict the merger event distributions with respect to various variables such as the component masses $M_1$, $M_2$, their ratio $q$ and the redshift $z$. 

The results that we obtained throughout this paper are critically sensitive to the precise prescriptions that we have chosen to work with. 
The most crucial elements in our model are the rate kernel $\tilde{R}_{i,j}$ and the IMF. For simplicity, we decided to adopt a model where the IMF has no dependence on metallicity.
As for the rate kernel, rather than specifying a particular model, as in \cite{Christian:2018mjv, Mouri:2002mc}, here we followed a phenomenological approach and varied the parameters of the kernel.
We also assumed for simplicity that the overall rate regulator $R_{11}$ is constant with time, although a more refined model might consider this parameter to vary as the cluster evolves. 
In addition, in this work we assumed a single type of GCs, although a more thorough analysis would consider a distribution of GCs with different properties. 
Another modification would consist of augmenting the coagulation equation with terms that arise from the formation of new BHs via channels other than BHs/NSs coalescences (such as stellar collapse of young stars).
Finally, BH spins are an additional important piece of information that we left outside the scope of this work (the relative spins orientations can e.g.\ shed light on the origin of the detected BHs \citep{Belczynski:2017gds, Bogomazov:2018prw, Gerosa:2017kvu, Fishbach:2017dwv, Baibhav:2020xdf, Wysocki:2018mpo, Sedda:2020vwo, Safarzadeh:2020mlb, Miller:2020zox}). Plus, by monitoring the BH spins we can rule out more parameter configurations and simulate more accurately GWA ejections. We leave such modifications to future studies.

We have demonstrated that any detected BH with masses between $50$ and $\sim100M_\odot$ (i.e. in the upper mass gap) can be easily explained as the result of SBH coagulation (in fact, it will be hard to explain a vanishing number of these if the dynamic channel is dominant). 
As for the lower mass gap ( $<5 M_\odot$), we demonstrated it is possible to detect events within that region if NSs are also considered as part of the coagulation process. 

When we considered the future O5 run, we were able to forecast excellent constraints on the parameters of both our coagulation kernel as well as the IMF. Moreover, our analysis suggests that future measurements may tell us (within 2-$\sigma$ CL) whether the BHs we detect are located in either dynamic environments (such as GCs) or in much less dense environments (such as the galactic field).

By tracking the merger rate history (Fig.~\ref{figure_7}) we showed the importance of including the delay time effect in any analysis of hierarchical mergers in evolving clusters. Even though this effect changes the history of the cluster in appreciably and non-trivially manner, our findings suggest that the current generation of GW detectors is only marginally sensitive to it. However, it is crucial to stress that if one attempts to calculate the interaction rate regulator $R_{11}$ from first principles (rather then normalizing it to achieve LIGO rate, as was done in this work), then the delay time effect must be considered.

As a byproduct of our analysis, we were able to calculate the probability to form an IMBH throughout the lifetime of the GC. Fig.~\ref{figure_4} shows that it is plausible to form an IMBH via hierarchical mergers of SBHs if the kernel of the GC diverges, which can happen for either positive $\gamma$ values or negative $\beta$ values. Ejections, mass loss and delay time lower considerably the likelihood to form IMBH through hierarchical mergers within Hubble time, but if most of the GCs contain many more BHs we can still expect to have roughly one IMBH form in each MW-like galaxy in the Universe if $\beta<-2$.

The tools developed in this work provide a sound basis for an analytical framework to study the properties of compact-object merger events detected in 
 GW observatories, which can be easily expanded to include more relevant physical effects and additional merger channels. As GW detection technology advances and more BH mergers are observed, such tools, complementary to N-body simulations, will be indispensable if we are to gain deeper insight regarding the sources of GWs in our Universe.

We thank Ofek Birnholtz and Sunny Itzhaki for useful insights. We also thank the anonymous referee for suggestions that helped improve the quality of the paper. JBM is supported by NSF grant AST-1813694 at Harvard and the Clay Fellowship at the Smithsonian Astrophysical Observatory. EDK is supported by a Faculty Fellowship from the Azrieli Foundation.

\section*{Data Availability}
The data underlying this article will be shared on reasonable request to the corresponding author.

\appendix
\section{Derivation for the relation between the probability for ejection and the recoil velocity}\label{Derivation for the relation between the probability for ejection and the recoil velocity}
Consider a cluster of total mass $M$ and a core radius $R$ with a Plummer density profile ~\citep{Plummer1911}, 
\begin{equation}
\rho\left(r\right)=\frac{3M}{4\pi R^3}\left(1+\frac{r^2}{R^2}\right)^{-5/2}.
\end{equation}
Now consider an object of mass $m$ with a recoil velocity $v_\mathrm{rec}$ at distance $r$ from the cluster's center. This object will eventually get ejected from the cluster only if its energy is positive
\begin{equation}\label{Energy}
E=\frac{1}{2}mv_\mathrm{rec}^2+m\Phi\left(r\right),
\end{equation}
where $\Phi\left(r\right)$ is the gravitational potential at distance $r$. For a Plummer density profile, the gravitational potential is
\begin{equation}
\Phi\left(r\right)=-\frac{GM}{\sqrt{r^2+R^2}}.
\end{equation}
Therefore, the condition $E>0$ is equivalent to
\begin{equation}
r>r_\mathrm{ej}\equiv R\sqrt{\left(\frac{2GM}{Rv_\mathrm{rec}^2}\right)^2-1}.
\end{equation}
Thus, we learn that given a recoil velocity $v_\mathrm{rec}$, ejection will happen only if $r>r_\mathrm{ej}$. Note that if $v_\mathrm{rec}^2>2GM/R$ then $r_\mathrm{ej}$ becomes imaginary, indicating that ejection will occur for each value of $r$, i.e. $P_\mathrm{ej}=1$ in that case. For $v_\mathrm{rec}^2<2GM/R$, we marginalize over $r$,
\begin{eqnarray}
\nonumber P_{\mathrm{ej}}&&=\frac{1}{M}\int_{0}^{\infty}\mathrm{d}^{3}r\rho\left(r\right)\mathcal{H}\left(E\right)=\frac{1}{M}\int_{r_\mathrm{ej}}^{\infty}\mathrm{d}^{3}r\rho\left(r\right)
\\&&=1-\left(\frac{R^2}{r_\mathrm{ej}^2}+1\right)^{-3/2}.
\end{eqnarray}
In conclusion,
\begin{equation}
P_{\mathrm{ej}}=\begin{cases}
\displaystyle{1-\left(1-\frac{v_\mathrm{rec}^4}{v_\mathrm{esc}^4}\right)^{3/2}} & v_\mathrm{rec}\leq v_\mathrm{esc}\\
1 & v_\mathrm{rec}\geq v_\mathrm{esc}
\end{cases},
\end{equation}
where $v_\mathrm{esc}^2\equiv 2GM/R$.

\section{Derivation for 2-body capture rate kernel}\label{Derivation for 2-body capture rate kernel}
Given the cross section $\tilde{\sigma}_{i,j}$ for the interaction of two masses $M_{i}$ and $M_{j}$, we can calculate the rate $\Gamma_{i,j}$ of such interactions
\begin{equation}\label{rate equation}
\Gamma_{i,j}=\int\mathrm{d}^{3}rn_{i}\left(\boldsymbol{r}\right)n_{j}\left(\boldsymbol{r}\right)\left\langle v_{\mathrm{rel}}\tilde{\sigma}_{i,j}\left(v_{\mathrm{rel}}\right)\right\rangle ,
\end{equation}
where $n_{i}\left(\boldsymbol{r}\right)$ is the number density of BHs with mass $M_{i}$ and $v_{\mathrm{rel}}$ is the relative velocity between the BHs.

The brackets denote averaging, which is defined as
\begin{equation}\label{velocity average}
\left\langle v_{\mathrm{rel}}\tilde{\sigma}_{ij}\left(v_{\mathrm{rel}}\right)\right\rangle =\int\mathrm{d}^{3}v_{\mathrm{rel}}v_{\mathrm{rel}}\tilde{\sigma}_{i,j}\left(v_{\mathrm{rel}}\right)\psi\left(v_{\mathrm{rel}}\right),
\end{equation}
where $\psi\left(v_{\mathrm{rel}}\right)$ is the relative-velocity distribution. If we assume that the velocity of BHs of mass $M_{i}$ distributes as a Maxwell-Boltzmann then
\begin{equation}
\psi\left(v_{i}\right)=\left(\frac{1}{2\pi\sigma_{i}^{2}}\right)^{3/2}\mathrm{exp}\left(-\frac{v_{i}^{2}}{2\sigma_{i}^{2}}\right),
\end{equation}
where $\sigma_{i}^{2}\propto M_{i}^{-1}$ is the dispersion velocity. In that case, $\psi\left(v_{\mathrm{rel}}\right)$ also distributes as Maxwell-Boltzmann but with a dispersion velocity
\begin{equation}\label{relative velocity dispersion}
\sigma_{\mathrm{rel}}^{2}=\sigma_{i}^{2}+\sigma_{j}^{2}\propto M_{i}^{-1}+M_{j}^{-1}=\sigma_{0}^{2}\frac{i+j}{i\cdot j}.
\end{equation}
where $\sigma_{0}$ is the dispersion velocity of BHs of mass $M_{0}$. We can then calculate the integral in Eq. (\ref{velocity average}) by using dimensional analysis:\footnote{Don't be confused by the double meaning of $\sigma: \tilde{\sigma}_{i,j}$ is the cross section whereas $\sigma_{\mathrm{rel}}$ is the relative dispersion velocity.}
\begin{equation}\label{velocity integral}
\int\mathrm{d}^{3}v_{\mathrm{rel}}v_{\mathrm{rel}}\tilde{\sigma}_{i,j}\left(v_{\mathrm{rel}}\right)\psi\left(v_{\mathrm{rel}}\right)=C\sigma_{\mathrm{rel}}\tilde{\sigma}_{i,j}\left(\sigma_{\mathrm{rel}}\right),
\end{equation}
where $C$ is a numerical constant which is determined from the velocity dependence of the cross section.

Mass segregation, driven by dynamical friction, will push heavier objects to the center of the core, giving them a steeper density profile than that of stars ~\citep{Mouri:2002mc, Sigurdsson:1994ju, Plummer1911}. For simplicity, we assume that all of the BHs sink to the core of the GC and take a flat profile over a core radius $R_{\mathrm{c}}$. This makes the volume integral 
\begin{equation}\label{volume integral}
\int\mathrm{d}^{3}rn_{i}\left(\boldsymbol{r}\right)n_{j}\left(\boldsymbol{r}\right)=\frac{3N_{i}N_{j}}{4\pi R_{\mathrm{c}}^{3}},
\end{equation}
where $N_{i}$ is the total number of BHs of mass $M_{i}$.

Plugging Eq. (\ref{velocity integral}) and (\ref{volume integral}) back into Eq. (\ref{rate equation}), we have
\begin{equation}\label{rate expression}
\Gamma_{i,j}=\frac{3CN_{i}N_{j}}{4\pi R_{\mathrm{c}}^{3}}\sigma_{\mathrm{rel}}\tilde{\sigma}_{i,j}\left(\sigma_{\mathrm{rel}}\right).
\end{equation}

We take the analytical expression for the cross section of 2-body gravitational capture ~\citep{QuinlanShapiro1989}
\begin{eqnarray}\label{2-body cross section}
\nonumber\tilde{\sigma}_{i,j}\left(v_{\mathrm{rel}}\right)&=&\pi\left(\frac{340\pi\mu}{3M_{\mathrm{tot}}}\right)^{2/7}\frac{M_{\mathrm{tot}}^{2}G^{2}}{v_{\mathrm{rel}}^{18/7}}
\\&\approx&16.83\frac{M_{0}^{2}G^{2}}{v_{\mathrm{rel}}^{18/7}}\left(i\cdot j\right)^{2/7}\left(i+j\right)^{10/7}.
\end{eqnarray}
For that kind of cross section, the constant $C$ turns out to be $C=2^{3/14}\pi^{-1/2}\Gamma\left(5/7\right)\approx0.84$. After plugging Eq. (\ref{relative velocity dispersion}) and Eq. (\ref{2-body cross section}) into (\ref{rate expression}), we get
\begin{equation}
\Gamma_{i,j}=5.27\frac{N_{i}N_{j}G^{2}M_{0}^{2}}{R_{\mathrm{c}}^{3}\sigma_{0}^{11/7}}\left(i\cdot j\right)^{15/14}\left(\frac{i+j}{2}\right)^{9/14},
\end{equation}
and from here we can get to an expression for the normalized rate
\begin{equation}
R_{i,j}=5.27\frac{G^{2}M_{0}^{2}}{R_{\mathrm{c}}^{3}\sigma_{0}^{11/7}}\left(i\cdot j\right)^{15/14}\left(\frac{i+j}{2}\right)^{9/14}.
\end{equation}

We can now identify the kernel $\tilde{R}_{i,j}$,\footnote{When King profile is used (instead of the flat profile we assumed in Eq. \ref{volume integral}) to account for mass segregation, the power index of the $\left(i\cdot j\right)$ term receives an additional contribution of 3/2 ~\citep{Mouri:2002mc, Sigurdsson:1994ju}.}
\begin{equation}
\tilde{R}_{i,j}=\left(i\cdot j\right)^{15/14}\left(\frac{i+j}{2}\right)^{9/14}.
\end{equation}
In addition, we have an expression for the normalization factor $R_{11}$,
\begin{eqnarray}\label{R11 expression}
R_{11}&=&5.27\frac{G^{2}M_{0}^{2}}{R_{\mathrm{c}}^{3}\sigma_{0}^{11/7}}
\\\nonumber&=&1.5\times10^{-18}\mathrm{yr}^{-1}\left[\frac{1\mathrm{km/s}}{\mathrm{\sigma_{0}}}\right]^{11/7}\left[\frac{1\mathrm{pc}}{R_{\mathrm{c}}}\right]^{3}\left[\frac{M_{0}}{\mathrm{1M_{\odot}}}\right]^{2}.
\end{eqnarray}
We can interpret Eq. (\ref{R11 expression}) in the following way: For a GC of typical radius 1pc, BHs of mass $1M_{\odot}$ which have dispersion velocity of 1~km/s are likely to interact every $0.66\times10^{18}$ years.

\section{Mass dependent $\lowercase{t_{d,\mathrm{min}}}$}\label{Mass dependent td,min}
The minimum delay time $t_{d,\mathrm{min}}$, defined in Eq.~\eqref{delay time distribution}, is the shortest time a binary must wait before it merges. For simplicity, in this work we assumed a global minimum delay time of $50\,\mathrm{Myears}$, regardless of the mass contents of the binary. In reality however, $t_{d,\mathrm{min}}$ may depend on the coalescing masses. This dependence complicates the coagulation equation quite a bit, but our framework is flexible enough to support it.

\citet{Fishbach:2021mhp} have found that the minimum delay time is more likely to be shorter for heavier binaries. With that statement in mind, we consider the simplest mass dependence model for $t_{d,\mathrm{min}}$ and assume it linearly decreases with the heavier mass of the binary,
\begin{eqnarray}\label{t,d,min(M1)}
\nonumber t_{d,\mathrm{min}}\left(M_i,M_j\right)&=&\frac{50M_\odot-M_1}{50M_\odot-15M_\odot}\left[t_{d,\mathrm{min}}\left(15M_\odot\right)-50\right]
\\&+&50\,\mathrm{Myears},
\end{eqnarray}
where $M_1=\max\left(M_i,M_j\right)$. Here, we fix $t_{d,\mathrm{min}}\left(50M_\odot\right)=50\,\mathrm{Myears}$ and we have only one degree of freedom, $t_{d,\mathrm{min}}\left(15M_\odot\right)$, which effectively controls the slope of $t_{d,\mathrm{min}}\left(M_1\right)$. Note that Eq.~\eqref{t,d,min(M1)} may yield negative values for very heavy binaries, in which case we set $t_{d,\mathrm{min}}$ to be the time resolution of the simulation.

Now, the delay time distribution for a given binary with masses $M_i$ and $M_j$ at time $t$ becomes
\begin{equation}
P\left(t_d;t,M_i,M_j\right)\propto\frac{1}{t_d}\mathcal{H}\left(t-t_d\right)\mathcal{H}\left[t_d-t_{d,\mathrm{min}}\left(M_i,M_j\right)\right],
\end{equation}
and the merger rate is
\begin{flalign}
\nonumber&\tilde\Gamma_{i,j}\left(t\right)=\int_0^t\Gamma_{i,j}\left(t-t_d\right)P\left(t_d;t,M_i,M_j\right)dt_d&
\\&\hspace{9mm}=R_{11}\tilde R_{i,j}\int_0^tN_i\left(t_d\right)N_j\left(t_d\right)P\left(t-t_d;t,M_i,M_j\right)dt_d.&
\end{flalign}

We study in Fig.~\ref{figure_7} how the total merger rate is changed due to the dependence of $t_{d,\mathrm{min}}$ on mass, for three different slopes, represented by the three blue curves. The solid curve corresponds to a zero slope, or a constant $t_{d,\mathrm{min}}$ of 50 Myears. In that scenario all the binaries are allowed to merge after 50 Myears, and merger rate evolves more slowly (compared to the $t_{d,\mathrm{min}}=0$ scenario). When a non-zero slope is introduced, the merger rate evolves even more slowly as can be seen by the dashed curve which corresponds to a slope of $7.14\,\mathrm{Myears}/M_\odot$ (or $t_{d,\mathrm{min}}\left(15M_\odot\right)=300\,\mathrm{Myears}$). The dotted curve corresponds to an extreme slope of $27.14\,\mathrm{Myears}/M_\odot$ (or $t_{d,\mathrm{min}}\left(15M_\odot\right)=1\,\mathrm{Gyears}$). Here we can see that the merger rate evolves so slowly that an inflection point emerges. This inflection point is the result of the formation of heavier BHs that merge more quickly as their $t_{d,\mathrm{min}}$ is shorter. 

\begin{figure}
\includegraphics[width=\columnwidth]{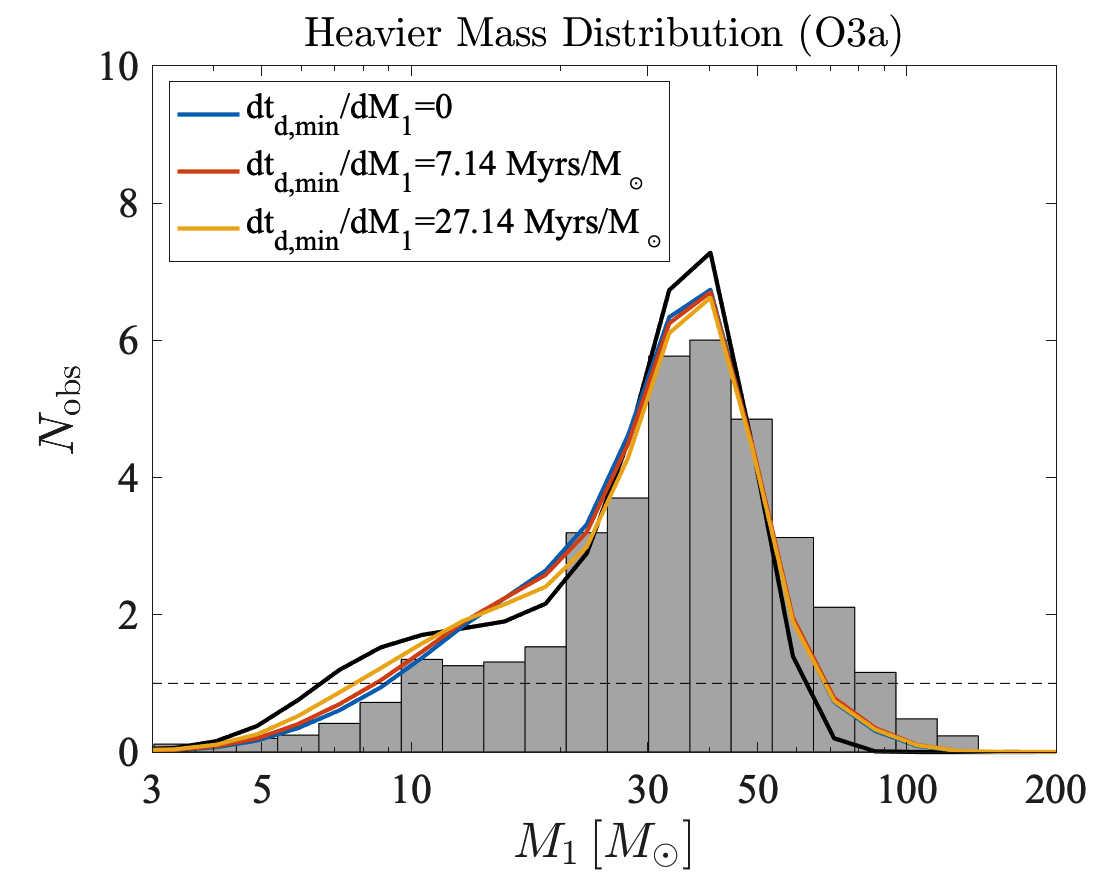}
\includegraphics[width=\columnwidth]{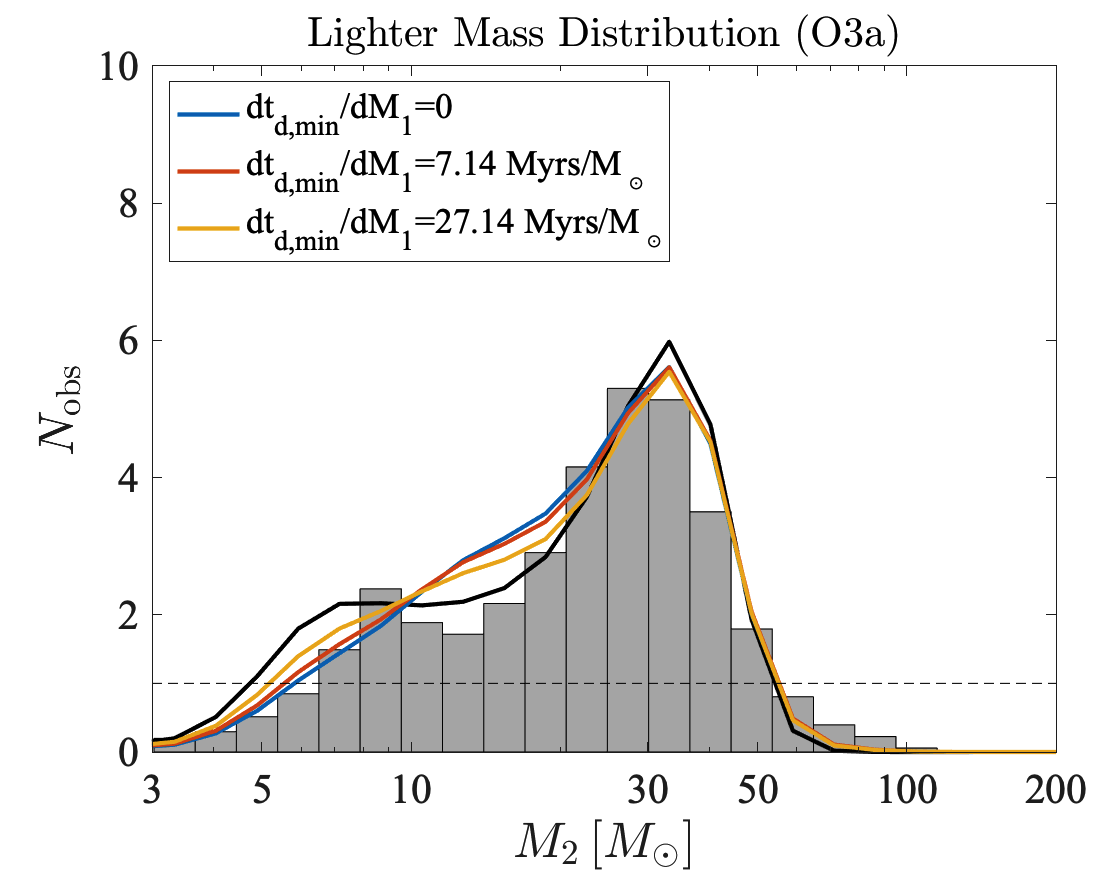}
\caption{The detected merger distribution with respect to $M_1$, $M_2$, predicted for O3a for different slopes of $t_{d,\mathrm{min}}$, assuming $\beta=2$, $\gamma=0$ and no ejections nor mass loss. The black curves correspond to $\beta_2=2$ static channel. The histograms correspond to O3a data. For each curve, the overall rate parameters $R_{11}$ and $R_0$ were chosen to match the total number of events in O3a (39).}
\label{figure_16}
\end{figure}

As we demonstrate in Fig.~\ref{figure_11}, the fact that the $\tilde\Gamma_\mathrm{tot}$ curve looks different does not necessarily mean that the resulting distribution of the detected merger events is different. In Fig.~\ref{figure_16} we examine how the $M_1$ and $M_2$ distribution change for different slopes of $t_{d,\mathrm{min}}$. Here we had to increase the value of $R_{11}$ as the slope increased in order to be consistent with the O3a run (39 events in total). Consequently, we see that there is almost no dependence of the events distributions on the slope of $t_{d,\mathrm{min}}$.

\section{Derivation for the GC age distribution}\label{Derivation for the GC age distribution}
In this appendix derive the GC age distribution $\psi\left(t\right)$ from theoretical considerations. We follow \cite{Fall:2001ti} and assume that the Globular Cluster Mass Function (GCMF) obeys the continuity equation,
\begin{equation}
\frac{\partial\psi}{\partial t}+\frac{\partial}{\partial M}\left(\dot{\psi}M\right)=0.
\end{equation}
The solution to this equation is
\begin{equation}\label{psi(M,t)}
\psi\left(M,t\right)=\psi\left(M_{0}\right)\frac{\partial M_{0}}{\partial M},
\end{equation}
where $\psi\left(M_{0}\right)$ is the GC initial mass function, and $M_{0}\left(M,t\right)$ is the initial mass of a cluster that has mass $M$ at a later time $t$.
\cite{Fall:2001ti} give the exact solution for $M\left(M_{0},t\right)$,
\begin{eqnarray}\label{M(M0,t)}
\nonumber M\left(M_{0},t\right)=&&\left\{ M_{0}-\mu_{ev}\int_{0}^{t}\mathrm{exp}\left[\nu_{sh}t'+S\left(t'\right)\right]dt'\right\}
\\&&\times\mathrm{exp}\left[-\nu_{sh}t-S\left(t\right)\right],
\end{eqnarray}
where $\mu_{ev}$ is a constant which is associated with the characteristics of two-body relaxation processes, $\nu_{sh}$ is a constant which is associated with the characteristics of gravitational shocks processes, and $S\left(t\right)=\int_{0}^{t}\nu_{se}\left(t'\right)dt'$ such that $\nu_{se}\left(t\right)$ is a function which is associated with the characteristics of stellar evolution processes.

We can invert Eq. (\ref{M(M0,t)}) to obtain $M_{0}\left(M,t\right)$,
\begin{eqnarray}
\nonumber M_{0}\left(M,t\right)=&&M\mathrm{exp}\left[\nu_{sh}t+S\left(t\right)\right]
\\&&+\mu_{ev}\int_{0}^{t}\mathrm{exp}\left[\nu_{sh}t'+S\left(t'\right)\right]dt'.
\end{eqnarray}
and from here,
\begin{equation}\label{dM0/dM}
\frac{\partial M_{0}}{\partial M}=\mathrm{exp}\left[\nu_{sh}t+S\left(t\right)\right].
\end{equation}
Plugging Eq. (\ref{dM0/dM}) back to Eq. (\ref{psi(M,t)}) gives
\begin{equation}
\psi\left(M_{0},t\right)=\psi\left(M_{0}\right)\mathrm{exp}\left[\nu_{sh}t+S\left(t\right)\right].
\end{equation}
From here, we can identify that 
\begin{equation}
\psi\left(t\right)\propto\mathrm{exp}\left[\nu_{sh}t+S\left(t\right)\right].
\end{equation}

\end{document}